\documentclass{aa}
\usepackage{graphicx}
\usepackage{natbib}
\bibpunct{(}{)}{;}{a}{}{,}    %to follow the A&A style

\usepackage{txfonts}
\newcommand{\beq}{\begin{equation}}
\newcommand{\eeq}{\end{equation}}
\newcommand{\bea}{\begin{eqnarray}}
\newcommand{\eea}{\end{eqnarray}}

\newcommand{\subscr}[1]{_\mathrm{#1}}

\newcommand{\url}[1]{{\tt #1}}

\newcommand{\lsim}{\raisebox{-0.6ex}{$\stackrel{{\displaystyle<}}{\sim}$}}
\newcommand{\gsim}{\raisebox{-0.6ex}{$\stackrel{{\displaystyle>}}{\sim}$}}
\def\gapp{\lower 3pt\hbox{${\buildrel > \over \sim}$}\ }
\def\lapp{\lower 3pt\hbox{${\buildrel < \over \sim}$}\ }

\begin{document}
 \title{Planet Formation in Binary Stars: The case of $\gamma$ Cep}

  \author{Wilhelm Kley\inst{1}
          \and
          Richard P. Nelson\inst{2}
          }

   \offprints{W. Kley}

   \institute{
        Institut f\"ur Astronomie und Astrophysik \&
        Kepler Center for Astro and Particle Physics,  
        Universit\"at T\"ubingen,\\
        Auf der Morgenstelle 10, D-72076 T\"ubingen, Germany\\
          \email{wilhelm.kley@uni-tuebingen.de}
         \and
      Astronomy Unit, School of Mathematical Sciences,
      Queen Mary, University of London, Mile End Road,
      London, E1 4NS, U.K. \\
            \email{R.P.Nelson@qmw.ac.uk}
             }
\date{Last revision: May 2008}
%\date{Received 30 May 2004 / Accepted 1 October 2004}
%%
%\maketitle

  \abstract
   {Over 30 planetary systems have been discovered to reside in binary stars.
   As some of the binary separations are smaller than 20 astronomical units (AU) the
   gravitational perturbation of the secondary star has a very strong influence on the
   planet formation process, as it truncates the protoplanetary disk, possibly
   shortens its lifetime, and stirs up the embedded planetesimals. Due to its small
   semi-major axis (18.5 AU) and relatively large eccentricity $e=0.35$ the binary star
    $\gamma$ Cephei represents a particularly challenging example worthy of study in
    greater detail.}
   {In the present study we model the orbital evolution and growth of embedded 
   protoplanetary cores of about 30 earth masses in the putative protoplanetary disk 
   surrounding the primary star in the $\gamma$ Cep system.}
   {We assume coplanarity of the disk, binary and planet and perform two-dimensional
    hydrodynamic simulations of embedded cores in a protoplanetary disk perturbed
    by a secondary companion. Before embedding the planet, the equilibrium structure
    of the disk for the observed binary parameters of $\gamma$ Cep is determined.
    We initiate the embedded planets in the disk on circular orbits with different
    initial distances from the primary. 
   }
   {The presence of the eccentric secondary star perturbs the disk periodically and
    generates strong spiral arms at periapse which propagate toward the disk centre.
    The disk perturbations then weaken as the secondary approaches apoapse. 
    The disk also becomes slightly eccentric 
    ($e_{disk} \approx 0.1-0.15$), and displays a slow retrograde precession in the inertial
    frame. Embedded cores interact with the eccentric disk, are periodically
    disturbed by the strong spiral shocks, and also by the eccentric binary. 
    We find that their
    eccentricity evolution depends primarily on the starting position in the disk.
    For all initial separations (from 2.5 to 3.5 AU) we find inward migration of
    the cores.
    For initial semi-major axes $a_p \gsim 2.7$, however, we find a strong increase 
    in the planetary
    eccentricity despite the presence of inward migration.
    Only cores which are initially far from the disk outer edge
    ($a_p \lsim 2.7$AU) have a bounded orbital
    eccentricity which converges, after mass accretion,
    roughly to the value of the planet observed
    in the $\gamma$ Cep system.
   }
   {Even though a close binary system such as $\gamma$ Cep still presents a challenge
    to planet formation theory, we have shown that under the condition that protoplanetary
    cores can form at around $2.5$ AU, it is possible to evolve and grow such a core 
    to form a planet with 
    final configuration similar to that observed.}
\keywords{planetary systems: formation -- accretion disks 
          hydrodynamics}
\maketitle
\markboth
{Kley \& Nelson: Planet Formation in $\gamma$ Cep}
{Kley \& Nelson: Planet Formation in $\gamma$ Cep}
\section{Introduction}
\label{sec:introduction}
Among the known 280 extrasolar planets around main sequence stars, about
33 are known to reside in binary systems with a wide range of orbital
separations \citep{2004A&A...417..353E, 2005Natur.436..230K,
2005A&A...440.1051M, 2007A&A...462..345D}. In all objects the planet orbits one
of the two stars, while the second star acts as an external  perturber to this
system. These are the so-called {\it S-Type} (stellar) orbits
\citep{1986A&A...167..379D}. A truly `circumbinary planet' orbiting around
a central binary star (a so called {\it P-Type} orbit) has not yet been detected.

In most of the systems the binary separation is large enough such that
the presence of a secondary may not have influenced the planet formation process
around one of the stars too strongly. The subsequent long term evolution of the planet in
a binary may be shaped by the Kozai effect even for wider binaries
\citep{2003ApJ...589..605W, 2005ApJ...627.1001T, 2007ApJ...670..820W}.
However, relatively short period binary orbits with binary separations of about 20 AU and less
have been found in a few systems, for example in Gl~86 
and notably in $\gamma$~Cephei. This raises
questions concerning the formation and stability of planets in such systems.
The reality of a planet in HD 188753 \citep{2005Natur.436..230K},
whose hypothetical existence caused some discussions about its origin in such a close
binary system \citep{2007ApJ...654..641J, 2005ApJ...633L.141P, 2005ApJ...635L..89P}, has
been put into question by newer observations \citep{2007A&A...466.1179E}. 
 
The mass-period and eccentricity-period distributions
of extrasolar planets in binaries show interesting trends
compared with planets around single stars, as presented
in \citet{2004A&A...417..353E}, and more recently in an updated version by
\citet{2007A&A...462..345D}.
It is argued that for short planetary periods ($ \lsim 40$ days) the distributions of
mass and eccentricities in binaries are significantly different
than for longer periods.
Surprisingly, the closer-in planets in binaries tend to have small eccentricities
very similar to or even smaller than in single stars.
Additionally, there seems to be a large fraction of high mass planets
with small orbital periods in binaries, that are not seen in single stars.
However, presently the statistics are too small to allow definite conclusions.
To improve on this deficiency in the amount of data, several new search programs tailored
in particular to planets
in binary systems have been initiated \citep{2005AAS...207.8402K, 2005AAS...20716602M}.
Considering that observationally more and more planets are found to
reside in binary stars, and may have orbital characteristics
different from planets orbiting single stars,
additional new calculations following the formation and evolution of such
systems are urgently required.

Since planets form within protoplanetary disks which are tidally perturbed by the presence
of a companion star one might expect consequences for the planet formation process.
The most prominent effects are a truncation of the disk and an increase
in the relative velocities of planetesimals.
The influence a secondary has on the growth process through collisions
from planetesimals to planetary embryos was studied first
through 3-body calculation by \citet{1974Icar...22..436H, 1978A&A....65..421H}
and later by \citet{1998Icar..132..196W}.
Their results implies that for a solar mass star the minimum distance
to an orbiting companion (periastron), such that planetesimals at 1AU can
grow to larger bodies, must always be larger than about 16 AU.
The critical semi-major axis scales only weakly with distance from the star.
In the last years these small number studies have been extended to
full N-body simulations including several hundred planetesimals. These
typically utilise a modern N-body integrator adapted for binary stars
\citep[e.g.][]{2002AJ....123.2884C}, allow for direct collisions and sticking
\citep{2004RMxAC..22...99L}, and also gas drag \citep{2005MSAIS...6..172T}.
For binary parameters such as $\alpha$ Cen terrestrial planet formation from an initial sea
of planetesimals can proceed similar to the Sun-Jupiter system \citep{2004RMxAC..22...99L}.

Another path of studying the formation of planets in binary stars was taken
by \citet{2000ApJ...537L..65N}, who studied the interaction of two binary stars each of
which is surrounded by its own circumstellar disk.
The two disks are both modelled hydrodynamically, and
during each binary orbit they are strongly perturbed at the time of periapse.
\citet{2000ApJ...537L..65N} argues that due to the periodic heating
of the disk planet formation becomes more unlikely.
On the one hand, the temperatures in the disk during periapse
become so high that condensation of solid material is reduced, while
on the other hand the increased
temperatures in the spiral arms hinders the onset of gravitational instability.
However, \citet{1998DPS....30.2309B,2006ApJ...641.1148B} has argued that
in some cases the presence of the secondary may even induce planet formation
through gravitational instability.

The dynamical evolution of Jovian protoplanets in circumbinary disks (P-type orbits)
has been investigated by \citet{2003MNRAS.345..233N} and \citet{2008A&A...482..333P}, 
studies which were extended recently to
include low mass protoplanets \citep{2007A&A...472..993P, 2008A&A...478..939P}. 

As of today there is only one rather preliminary
study \citep{2000IAUS..200P.211K}
dealing directly with the dynamical evolution of embedded protoplanets in
external binary stellar systems. In that study a Jupiter mass planet was evolved
in a protoplanetary disk for a range of binary separations $a_{bin}=50-100$AU
and a given binary eccentricity $e_{bin} =0.5$. In the binary
environment the planet experienced a faster
growth and migration with respect to the unperturbed case. However,
the simulations were rather unrealistic in the sense that the planet was not
grown in the disk from an embryo to its final size, but rather
placed `ready made' into an otherwise unperturbed disk.

In this contribution we would like to improve on these models, and
in a future publication we will present our results of
new simulations following
the evolution and growth of planetesimals in the protoplanetary disk
in a binary system. In this present study we consider the particular system
$\gamma$ Cep because this is one of the tighter systems, and thus places
rather strong constraints on the possible planet formation scenarios.
For these reasons, this system has attracted much attention
in past years. Several studies looked at the stability and/or the
possibility of (additional) habitable planets in the system
\citep{2004RMxAC..21..222D, 2004MSAIS...5..127T, 
2006ApJ...644..543H, 2006MNRAS.368.1599V}.
Finally, the possibility to grow protoplanets from very small
planetesimals in $\gamma$ Cep has been studied by  
\citet{2004A&A...427.1097T}, who placed km-sized planetesimals
in a protoplanetary disk including gas drag and gravitational
disturbances from the secondary and analysed the relative impact velocity.
For a sufficient amount of gas present, the aerodynamic drag between
gas and particle leads to periastron alignment such that that
runaway growth is possible inside of 2.5 AU from the primary star.
A more extended study of this process for various dynamical parameter
of the binary has been performed by \citet{2006Icar..183..193T}.
Recently \citet{2007arXiv0705.3421K} and \citet{2008arXiv0802.0927P} 
improved again on these models
by considering the evolution of a swarm of planetesimals embedded
in a hydrodynamically evolving disk.

In the present work we focus on $\gamma$ Cep and we assume,
along the work of \citet{2004A&A...427.1097T}, that protoplanetary 
cores of a few earth masses have successfully formed inside the disk and
follow the subsequent dynamical evolution and mass growth.
We focus here on the later phase of gas accretion only and will present
additional new results on planetesimal formation in a forthcoming paper
\citep[see also][]{2007arXiv0705.3421K}.
We perform full numerical, hydrodynamical evolutions of protoplanetary disks
for binary parameters typical for $\gamma$ Cep, analyse the equilibrium configuration
of the disk, and study the subsequent evolution of embedded protoplanets.

This paper is organised as follows: in Sect.~\ref{sec:model-setup} we
describe the model set--up, and in Sect.~\ref{no-planet} we present results
of disk evolution due to the binary but in the absence of the embedded
planet. In Sect.~\ref{with-planet} we present results of simulations
with planets embedded in a protoplanetary disk perturbed by a
binary companion. Finally, we draw our conclusions in Sect.~\ref{conclusion}.

\begin{table}
\caption{Orbital data of the binary $\gamma$ Cep and its planet as given
by \citet{2003ApJ...599.1383H}, which have been used for the simulations.
}
\label{table:1}
\centering
\begin{tabular}{c c c c c c c}
\hline\hline
  $M_1$       &  $M_2$      &  $a$  & $e$ & $M_p$        & $a_p$  & $e_p$\\
 $~$[$M_\odot$]  & [$M_\odot$] &  [AU] &     &  [$M_{Jup}$] & [AU]   &    \\
\hline
  1.59   & 0.378  & 18.5  & 0.36 &  1.70 &  2.13 & 0.20 \\
\hline
\end{tabular}
\end{table}
\section{Model Setup}
\label{sec:model-setup}
In modelling the putative early evolution of the
$\gamma$ Cep system, we assume that the protoplanetary disk
is in an S-type configuration and orbits around the more massive
primary star. To clarify the primary effects of planetary growth
and evolution in binaries, we avoid complex
three-dimensional effects and assume that the 
whole system is coplanar, i.e. the orbits
of the disk, the binary and the planet are all lying in one plane
and the angular momenta are all aligned.
To further simplify the calculations we treat the disk in a two-dimensional
approximation, neglect any vertical extension and work with the
vertically integrated equations.
This approximation is often made for accretion disks
and applies whenever the vertical thickness $H$ (pressure scale height)
of the disk is small compared to the radial distance $r$ from the central star,
i.e. $h\equiv H/r << 1$.

We use cylindrical coordinates ($r, \varphi, z$), where the
the origin of the coordinate system lies in the centre of the primary star at
$r=0, z=0$.
The evolution of all objects (stars, disk and planet) takes place in
the $z=0$ plane.
As the stars orbit each other, the coordinate system is non-inertial
and the indirect terms have to be included in the equations of motion.

Accretion disks are most-likely driven by magneto-hydrodynamical turbulence
which we model here in a simplified manner through a viscous stress tensor model.
For the kinematic viscosity coefficient $\nu$ we assume a standard 
$\alpha$-prescription with $\nu = \alpha c_s H$, where $\alpha$
is a constant, and $c_s$ is the local sound speed given by $c_s(r) = h \, u_{kep}(r)$
where $u_{kep}(r)$ is the Keplerian velocity of the accretion disk around
the primary star.

The basic hydrodynamic equations (mass and momentum conservation)
describing the time evolution of such a viscous
two-dimensional disk with embedded planets have been stated frequently and are
not repeated here \citep[see][]{1999MNRAS.303..696K}.

The dynamical effects of the companion star will lead to a disk structure
which is different from that which exists around a single star.
To account for this restructuring of the disk we first model the disk
evolution in the binary system alone without any embedded planet.
As will be shown below this initialisation phase takes about 100
binary orbits. Only after this time the disk has relaxed (on average)
to its new equilibrium and then the planets are inserted into the models.

\subsection{Initial Conditions}
The two-dimensional ($r-\varphi$) computational domain consists of a
complete ring ($\varphi\subscr{min}=0$, $\varphi\subscr{max}=2\pi$)
of the protoplanetary disk that radially extends from
$r\subscr{min}=0.50$AU and $r\subscr{max}=8.0$AU.
The computational domain is covered by 300 $\times$ 300 ($N_r \times N_\varphi$)
grid cells, that are spaced equidistant in both radius and azimuth.

The initial hydrodynamic structure of the disk (density, temperature, velocity)
is axisymmetric with respect to the location of the primary star.
The surface density has the profile $\Sigma(r) = \Sigma_0 r^{-1/2}$
over the entire radial domain, where
$\Sigma_0$ is chosen such that the total mass in the computational domain
(within $r_{min}$ and  $r_{max}$) equals $1.75 \times 10^{-3} M_\odot$ which
implies $\Sigma_0 = 1.89 \times 10^{-5} M_\odot$/AU$^2$.
The initial velocity is pure Keplerian rotation ($u_r=0,
u_\varphi = \Omega_K r = (G M_*/r)^{1/2}$), and 
the temperature stratification is always given by
$T(r) \propto r^{-1}$ which follows from the assumed
constant vertical height $h = H/r =0.05$.
For these locally isothermal models the temperature profile is left unchanged
at its initial state throughout the computations, i.e. no energy
equation is solved.

\subsection{Boundary conditions}
The boundary conditions are chosen such that material may escape through the radial
boundaries. At the outer boundary ($r_{max}$) we impose a so called zero-gradient
outflow condition, where the velocity is extrapolated when directed outward
and set to zero when directed inward in the last active gridcell inside $r_{max}$.
For all other variables zero gradients are assumed.
During periastron when large spiral arms may extend beyond $r_{max}$ this condition will
allow material to leave the system freely and will help to reduce numerical artifacts.
At the inner boundary we set a viscous outflow condition where the material may
flow through $r_{min}$ with the local (azimuthally averaged) viscous inflow referring
to an accretion disk in equilibrium,
\beq
     u_r (r_{min}) = - \, \frac{3}{2} \frac{\bar{\nu}}{r_{min}} ,
\eeq
where $\bar{\nu}$ denotes the azimuthally averaged viscosity at $r_{min}$.
We experimented with an outflow inner boundary similar to that used at $r_{max}$
but found that in such a case the induced ellipticity of the disk
(by the eccentric binary) will cause strong outflow of disk material in its
periastron phase which in turn produces a large eccentric inner hole of
the disk which is a numerical artefact . Our chosen inner boundary allows
for accretion onto the star and additionally mimics the expected nearly circular
flow in the inner regions of the disk close to the primary.
No matter is allowed to flow back into the system and the mass of the disk will slowly
decline.

\subsection{Physical parameters}
In the majority of our our numerical simulations we choose to model a specific
system where the orbital elements of the binary have been chosen to match the system 
$\gamma$ Cep quite closely.
The data for this system have been taken from \citet{2003ApJ...599.1383H}
which do not include the most
recent observational improvements \citep{2007A&A...462..777N}.
These newest refinements primarily concern the mass of the
primary and should not alter our main conclusions at all.
The present studies are supposed to demonstrate
the principle physical effects occurring in close binaries rather than trying to
achieve a perfect match with all the observations of this system.
For the runs we choose a binary with
$M_1 = 1.59 M_\odot$, $M_2 = 0.38 M_\odot$,
$a_{bin} = 18.5$~AU and $e_{bin} = 0.36$ (see Table~\ref{table:1}),
which translates into a binary period of $P = 56.7$~yrs. 
For the viscosity we use a typical value for the
effective viscosity in a protoplanetary disk, i.e. $\alpha = 5 \times 10^{-3}$,
and a locally isothermal disk with $h=H/r = 0.05$. 

As mentioned above, the initial total disk mass in the system is 
$1.75 \times 10^{-3} M_\odot$.
Due to the presence of the secondary the disk will be truncated to a smaller
outer radius and the radial density stratification will adjust to a new
equilibrium configuration.
At the same time mass will be lost from the system due
to possible outflow through the (open) boundaries.
To ensure a somewhat uniform initial configuration for the planets
we rescale the disk mass to a given mass before inserting the planets.

\subsection{A few remarks on numerical issues}
\label{subsec:numerical}
To substantiate our results we use two different codes for our
calculations, {\tt RH2D}
\citep{1999MNRAS.303..696K, 1989A&A...208...98K}  and {\tt
NIRVANA} \citep{2000MNRAS.318...18N, 1997ZiegYork}, both
of which have been utilised frequently in embedded planet problems.
The numerical method used in both codes is a staggered
mesh, spatially second order finite difference method based, where
advection is based on the monotonic transport algorithm
\citep{1977JCoPh..23..276V}.  Due to operator-splitting the codes are
semi-second order in time.  The computational details of {\tt RH2D}
which can be used in different coordinate systems have been described
in general in \citet{1989A&A...208...98K}, and specifically for planet
calculations in \citet{1999MNRAS.303..696K}.  The algorithmic details of the {\tt
NIRVANA} code have been described in \citet{1998CoPhC.109..111Z}.
In all of our simulations we utilise a non-rotating coordinate 
frame centred on the primary.

In calculating the gravitational effect of the planet we use a
smoothed potential of the form
\beq
    \Phi_P  =  - \frac{ G M\subscr{p}}{\sqrt{s^2 + \epsilon^2}}
\eeq
where $s$ is the distance from the planet to the (grid-)point
under consideration.
For the smoothing length of the potential we choose 
$\epsilon = 0.6 R\subscr{Hill}$, where 
\beq
  R\subscr{Hill} = r_P \left(\frac{q}{3}\right)^{(1/3)} .
\eeq
Here $r_P$ denotes the momentary distance of the planet from the star,
i.e. the size of the Hill radius scales with the distance from the star
and not the semi-major axis. This distinction is important for calculating
the torques in the case of eccentric orbits. 

The smoothing length $\epsilon$ is proportional to
the local scale height $H$ of the disk to account for the vertical
extent of real disks. The chosen value of $0.6 R\subscr{Hill}$ gives a
reasonably good agreement between the linear torques of a three-dimensional
disk and the two-dimensional flat-disk approximation used here
\citep{2002ApJ...565.1257T}.
In a recent study of the dynamical evolution of eccentric low mass planets
we have found a good agreement for the migration and eccentricity
damping rate in 2D and 3D disks for this value of $\epsilon$
\citep{2007A&A...473..329C}. 

To model possible mass accretion onto the protoplanet we take out mass within
a given radius $r_{acc}$ around the planet at a prescribed rate.
Whenever the centre of a gridcell is closer than
$r_{acc} = 1/2 R_{Hill}$ we reduce the density in that cell by a factor
$1 - f_{acc} \, \Delta t$, where $\Delta t$ the the actual time--step
of the computation and $f_{acc}$ is a model dependent reduction factor.
The rate is doubled whenever the distance is smaller than 1/4 $r_{Hill}$.
For details see \citet{1999MNRAS.303..696K}. 

The viscous terms, including all necessary tensor components,
are treated explicitly.
To ensure stability in the outer parts of the disk where we expect stronger
gradients in the flow due to the perturbations by the secondary,
an artificial bulk viscosity has been added, with a
coefficient $C\subscr{art} = 1.0$. For the detailed formulation
of the viscosity related issues and tests, see \citet{1999MNRAS.303..696K}.

To avoid excessively low densities in the outer part of the computational domain
we have found it preferable to work with a density floor, where the density
cannot fall below a specified minimum value $\Sigma\subscr{min}$.
For our purpose we use a value of $\Sigma\subscr{min} = 10^{-6}$ in dimensionless units,
where the initial density is of ${\cal{O}}(1)$ at $r=1.0$.

When calculating the gravitational forces of the disk acting on the planet
numerically for a given grid resolution a cutoff radius $r_t$ is necessary in
addition to the smoothing of the potential. This cutoff reduces the contribution
of material in the direct vicinity of the planet which lies
inside the planetary Roche-lobe. This material is bound to the planet and typically
cannot exert any forces on it. Here we use a cutoff of $r_t = 0.8 R_{Hill}$
for most runs
which is smoothed at $r_t$ using a Fermi-function with a width of $0.1 r_t$.
The role of the cutoff radius $r_t$ in estimating the torque on the planet
has been analysed in more detail by \citet{2008A&A...483..325C}. For the range
$0.6 R_{Hill} \leq r_t \leq 1.0 R_{Hill}$ the obtained results agree very well with
each other, see also Sect.~\ref{subsec:exploratory} below.

The planet evolution and the orbital motion of the binary proceed
on time scales much longer than the typical dynamical time scale
of the disk, and consequently very many time--steps need to be calculated
just in one simulation. To allow for parameter studies we have increased
the performance of the runs by implementing the {\tt FARGO}-algorithm in
our codes, which is particularly designed to model differentially
rotating flows efficiently \citep{2000A&AS..141..165M}. 
For our chosen radial range and grid resolution we find a speed-up
factor of about 7.5 over the standard case. Then, applying a Courant number of 0.75,
still about 160,000 time--steps are required for only 10 binary orbits for our
radial range and a standard resolution of $300\times300$.
We checked the accuracy of the {\tt FARGO}
implementation against several test cases.
\section{Disk evolution without an embedded planet}
\label{no-planet}
When studying the formation of planets in a protoplanetary disk in the
presence of a secondary star it is necessary and illustrative to
first investigate the structure and evolution of the
accretion disk subject to the influence of the binary system only.
Hence, in this section we follow the evolution of a disk around the primary
which is perturbed by the secondary {\it without} an embedded planet.
As the disk evolves slowly from its initial axisymmetric state into
a new quasi-equilibrium which is quite different from the unperturbed
initial state, we have first to relax the system 
into this new equilibrium, before adding a planetary embryo at a later time.
In the following we describe in more detail this restructuring process of the
disk and the change in the binary elements due the finite disk mass.

\begin{figure}
 \centering
 \includegraphics[width=0.9\linewidth]{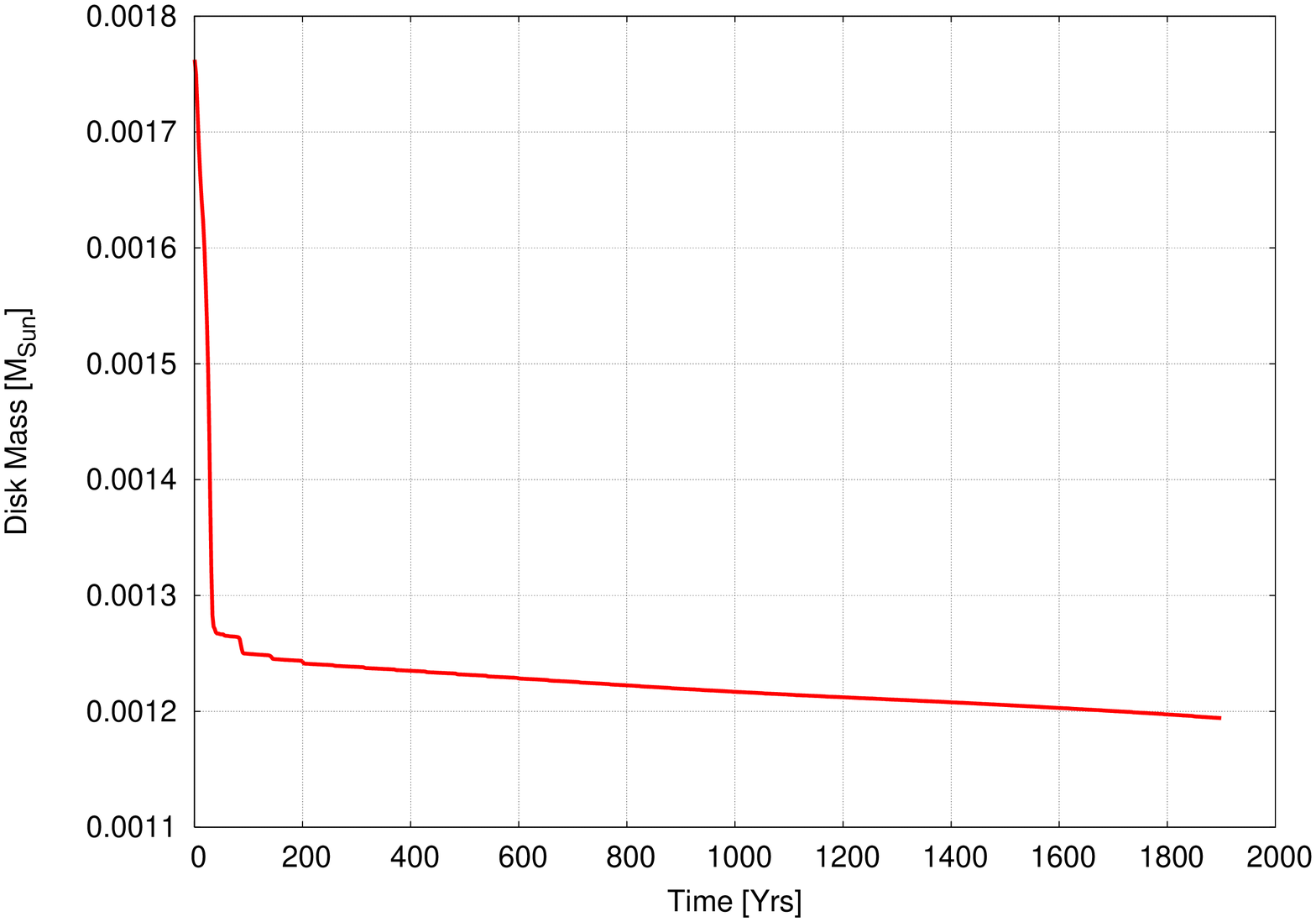}
 \caption{Time evolution of the total disk mass in units of the 
   $M_\odot$ vs. time in yrs for no embedded planet. 
  Note that one binary orbit refers to about 57 yrs.
   \label{fig:md-h03f}
   }
\end{figure}

\begin{figure}[ht]
\def\capfrac{1}
\resizebox{0.47\linewidth}{!}{%
\includegraphics{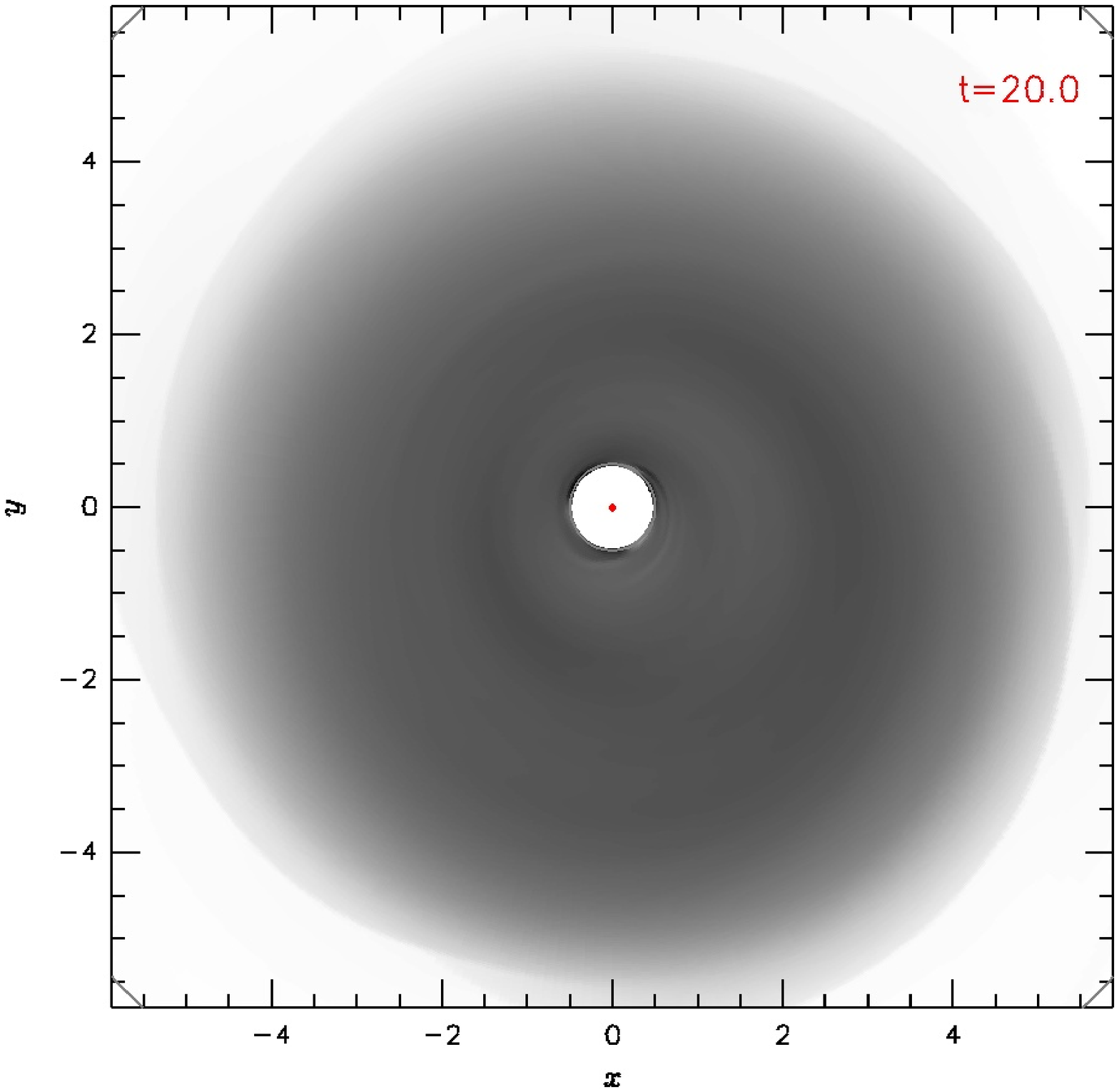}}
\resizebox{0.47\linewidth}{!}{%
\includegraphics{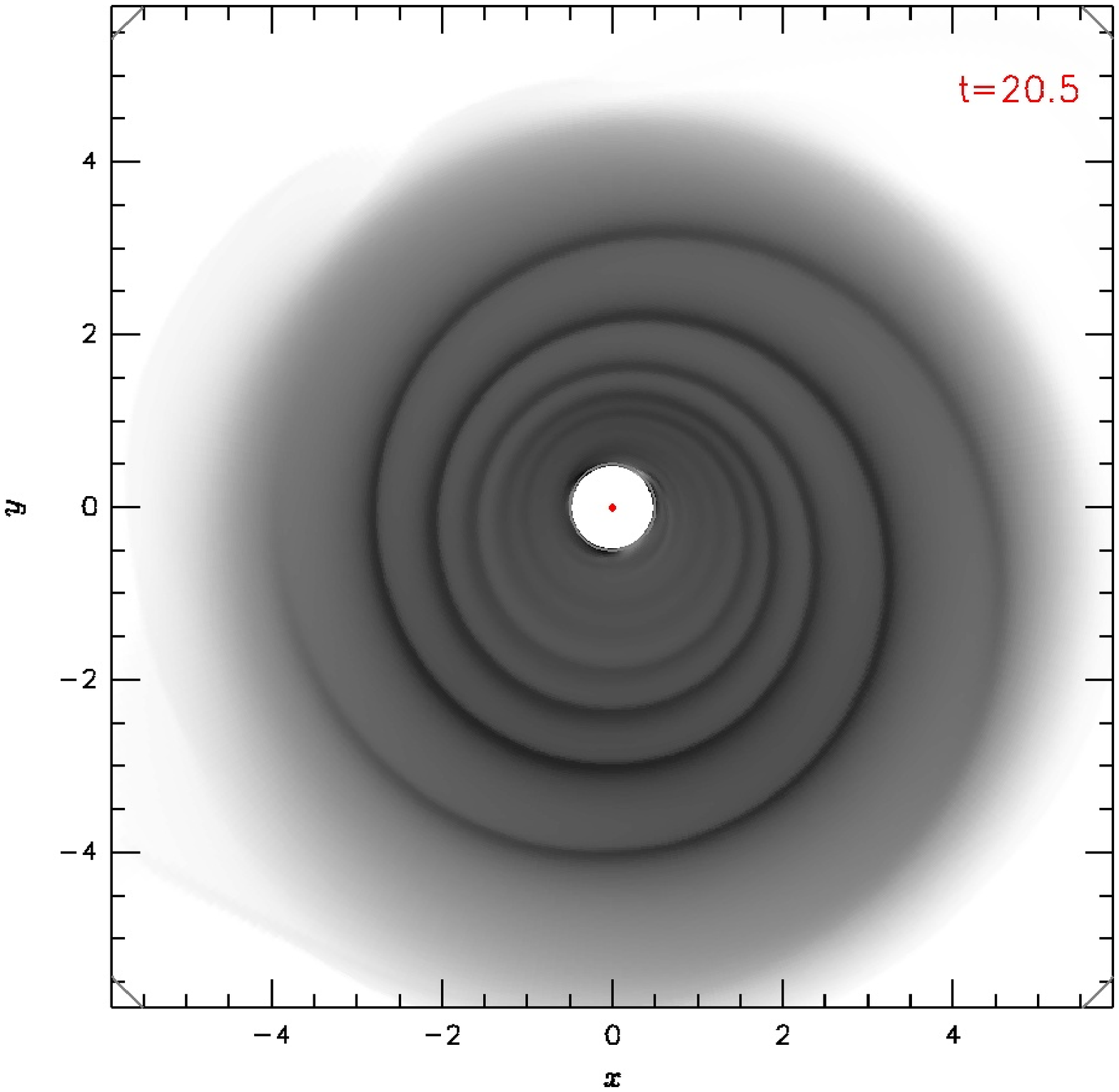}}
  \caption{
Grayscale plot of the two-dimensional density distribution
of the circumstellar disk around the primary at two different orbital
phases of the binary.
{\bf Left} shortly after apocentre at about 20 binary orbits, and
{\bf Right} shortly after closest approach (pericentre).
  }
   \label{fig:h03-xy}
\end{figure}

\begin{figure}[ht]
\def\capfrac{1}
\resizebox{0.90\linewidth}{!}{%
\includegraphics{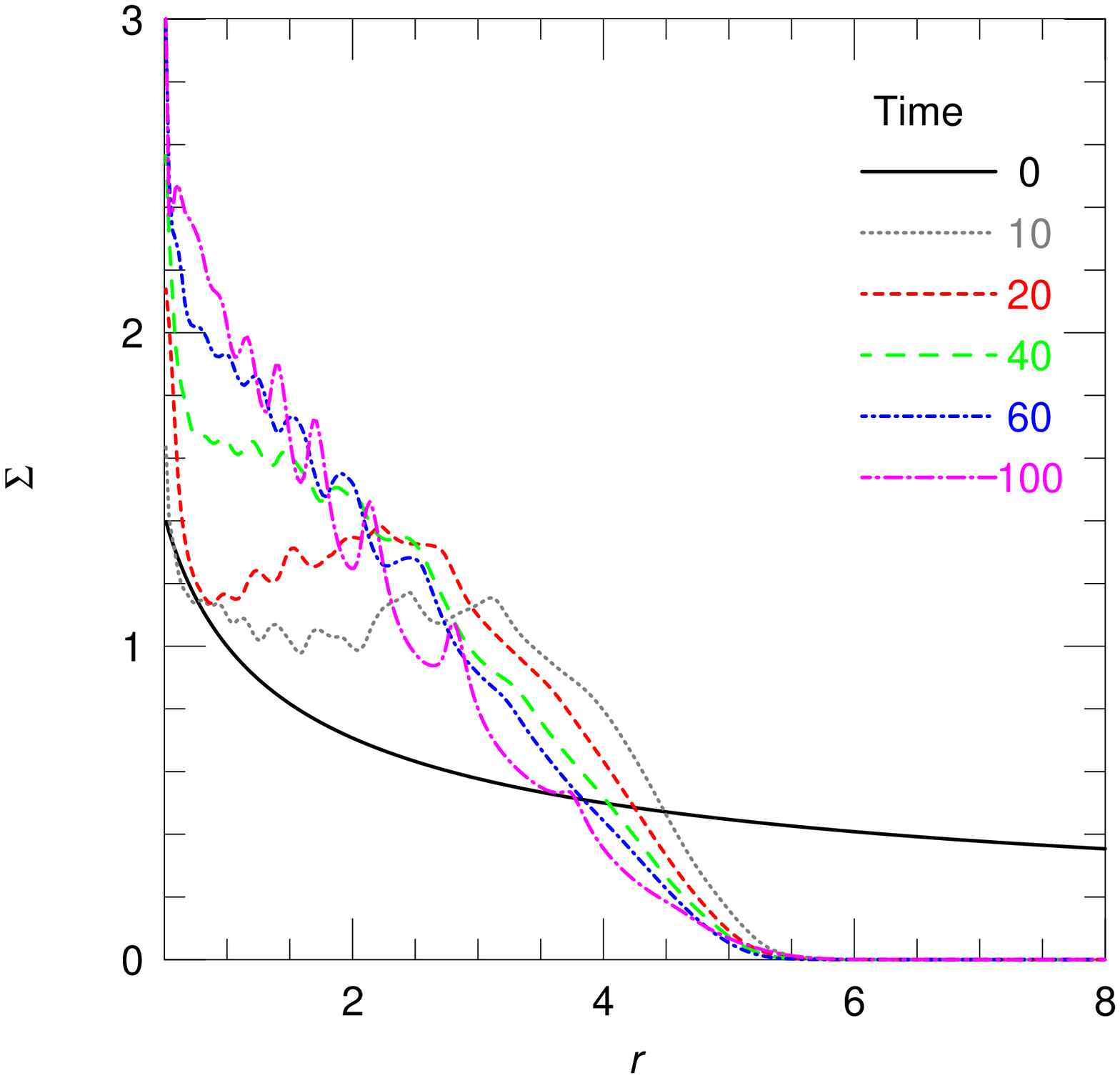}} \\
\resizebox{0.90\linewidth}{!}{%
\includegraphics{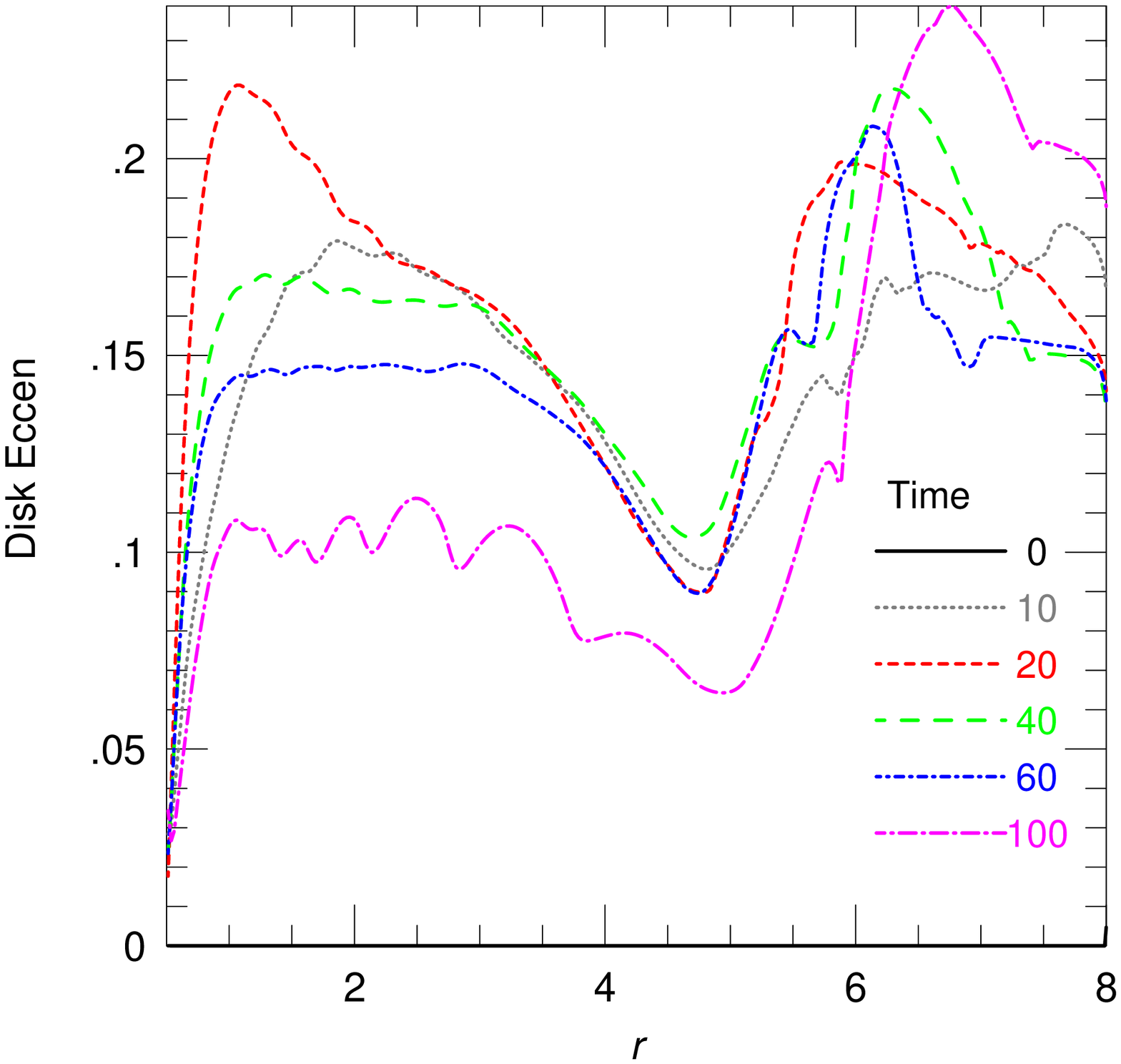}}
  \caption{
Radial dependence of the azimuthally averaged surface density ({\bf Top})
and the eccentricity ({\bf Bottom}) of the circumstellar disk around
the primary in the presence of the secondary at different times.
Time is given units of the binary orbit, radial distance
in AU, and the density in dimensionless units.
  }
   \label{fig:sig-ecc-h03}
\end{figure}

\subsection{The structure of the disk}
The presence of an eccentric secondary star leads to a strong periodic
disturbance of the disk whenever it is at periastron. Then two strong
tidal spiral arms (shock waves) are excited in the disk which wind themselves all
the way down to the centre.
At the outer edge of the disk the perturbation by the companion star
carries material
beyond the outer boundary of the computational domain which is then lost
from the system.  During the very first periapse phase
about 30\% of the initial disk mass is lost from the system, and very
little during the following passages, as show in Fig.~\ref{fig:md-h03f}.
In between the
periapses the disk settles down and becomes more circular again.
This effect is illustrated in the Fig.~\ref{fig:h03-xy}
where we display the surface density $\Sigma$ of the disk in gray scale
at 2 different times in the early evolution of the disk. In the left panel
the nearly circular state at the binary apoastron is displayed while
the right panel shows the situation just after periastron.
These findings are very similarly to those found by 
\citet{2000ApJ...537L..65N} in his study of the binary system L1551.

Already the very first close approaches with the binary lead to a clear truncation of
the disk at around $5$ AU, as is visible in left 
panel of Fig.~\ref{fig:sig-ecc-h03}
for the curve at $t=10$ binary orbits. Slowly the whole disk structure
rearranges and equilibrates at around $t=50$ where it shows a much steeper
(nearly linear) density slope than in the initial state.
The time scale for this equilibration process depends on the magnitude
of the disk viscosity and shock dissipation.
For relatively cool disks and small viscosities we expect that
the equilibrium profile is primarily determined by the same processes, i.e.
gravitational action of the
binary and shock dissipation.
The eccentricity of the disk in the final state of the disk varies approximately
between 0.1 and 0.16 depending on the position of the binary in its orbit
as shown in the bottom panel of Fig.~\ref{fig:sig-ecc-h03}.
In the outer parts of the disk beyond $r \approx 5$ AU, where the
density is very low, the disk eccentricity is measured to be
larger due to the proximity to the
secondary star.

\begin{figure}[ht]
\def\capfrac{1}
\resizebox{0.49\linewidth}{!}{%
\includegraphics{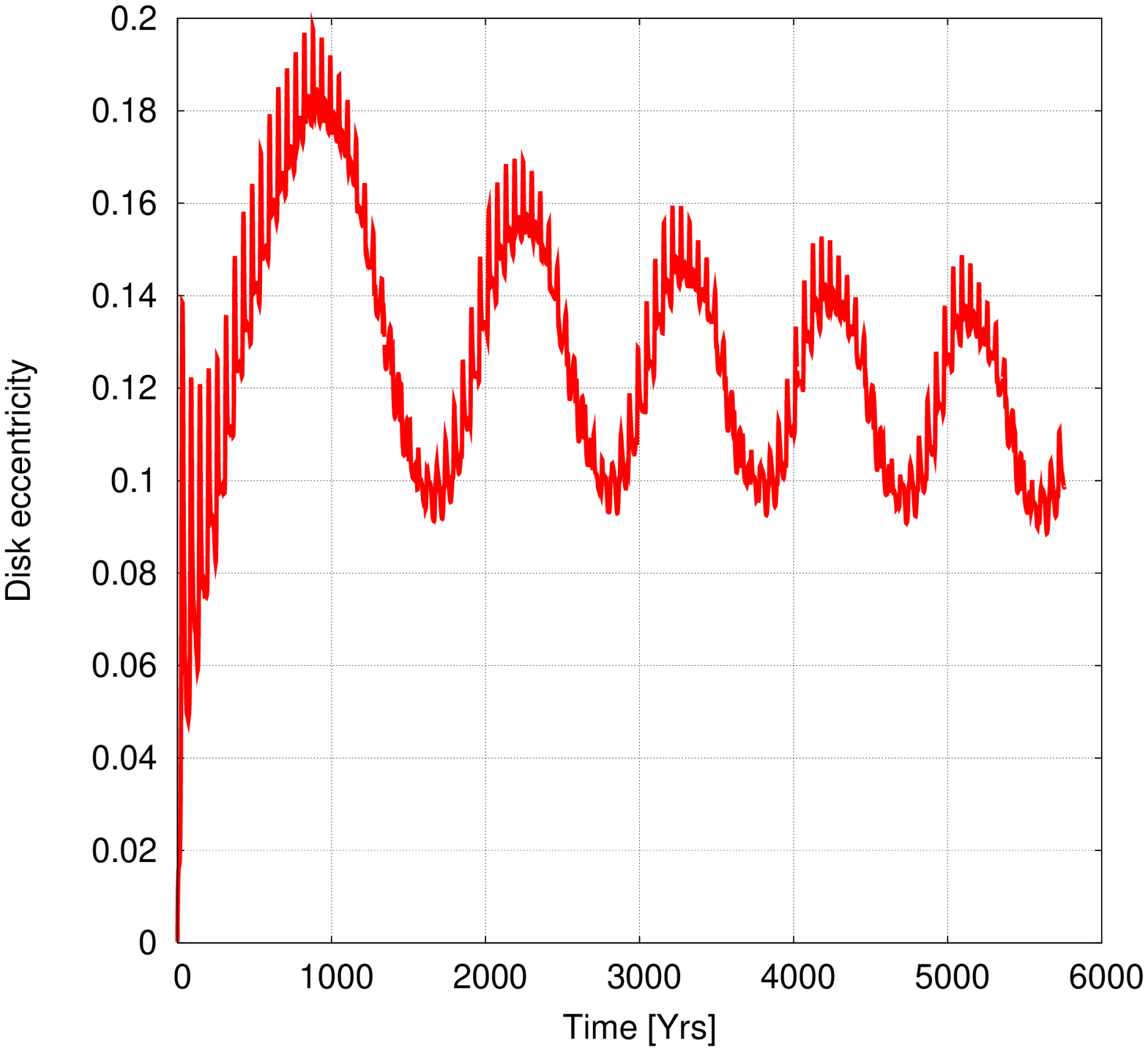}}
\resizebox{0.49\linewidth}{!}{%
\includegraphics{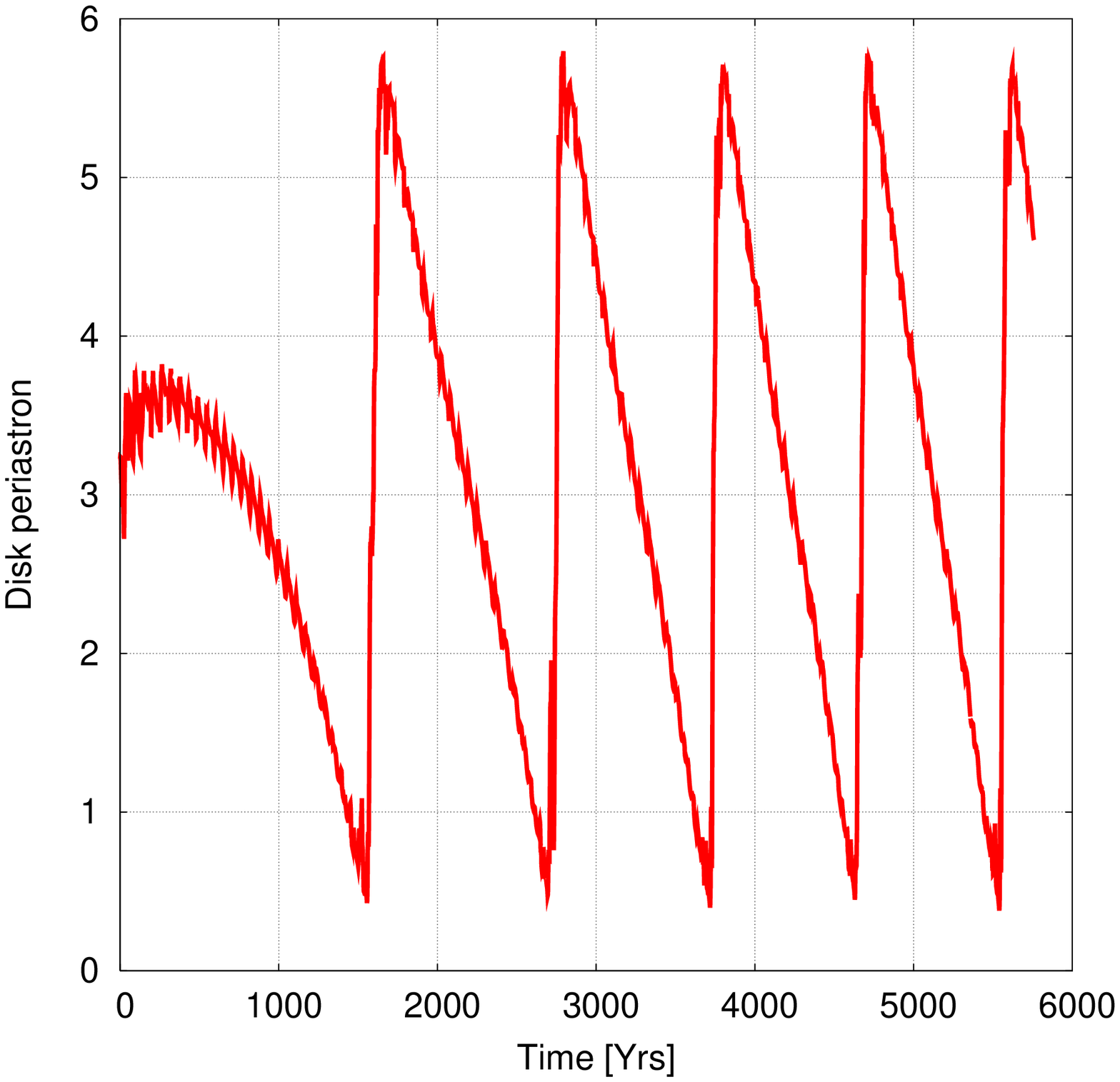}}
  \caption{
The evolution of the global (density weighted) averaged disk eccentricity ({\bf left}) and
the position angle of the disk's periapse ({\bf right}).
  }
   \label{fig:ecc-om-h03}
\end{figure}

The overall (mass averaged) eccentricity of the disk displays oscillations
on longer time scales of about 17 binary orbits around the mean value of 0.12
(left panel of  Fig.~\ref{fig:ecc-om-h03}). 
The disk eccentricity $e_{disk}(r)$ in that plot has been obtained by calculating the
eccentricity of each disk element, as if in a two body motion with the
primary star, and then averaged azimuthally over the respective annulus.
The superposed spikes are caused by the individual periapses of each orbit,
during which the whole disk is perturbed by the strong spiral arms. 
At the same time the disk as a whole precesses, as is shown in the right
panel of Fig.~\ref{fig:ecc-om-h03}. This coherent slow
precession is {\it retrograde} with a pattern speed much smaller than the
orbital period of the disk material around the star.
This interesting global disk phenomenon appears to be related to the
induced precession of accretion disks in close binary systems such as cataclysmic
variable stars which occurs even in the case of circular binaries
\citep{1991ApJ...381..259L}. Very similar results have recently been
found by \citet{2008arXiv0802.0927P}, see also \citet{2007arXiv0705.3421K}.
In test simulations using circular and
non-circular binaries we have also verified that eccentric disks occur
in the circular case as well, and in recent simulations \citet{2008arXiv0802.0927P}
have also demonstrated (for $\gamma$ Cep parameter) that in the limit of
circular orbits the mechanism that
causes the instability is indeed the 3:1 eccentric Lindblad resonance 
\citep{1991ApJ...381..259L}. In a related study we have analysed in detail
the influence of numerical parameter on the occurrence and properties of eccentric disks
in binaries and we refer the reader to that paper 
for more details (Kley, Papaloizou \& Ogilvie to be submitted). For the highly
truncated disk in this study we expect the averaged eccentricity to be relatively robust
against numerical variations. Test simulations of this initial equilibration
phase of the disk with different grid resolutions resulted in very similar values
of the disk eccentricity. For the typical values of disk viscosity and temperature
used by us we expect these values for the disk eccentricity, an expectation
which is confirmed by \citet{2008arXiv0802.0927P} who find find very similar results.
Concerning the evolution of embedded protoplanets in these
disks it appears that the effect of the periodic disturbance of the binary may
play a more important role than the eccentricity of the disk.

The direction of the precession (prograde or retrograde with respect to the
disk and binary motion) is influenced
by pressure effects in the disk. For our chosen value of $H/r =0.05$ we
do indeed expect a retrograde precession.
Similar behaviour has also been demonstrated
for disks with free eccentricity \citep{2005A&A...432..757P}.

\subsection{The orbital elements of the binary}
In the previous section we have seen that the gravitational
torques of the binary lead to truncation of the disk
and re-arrangement of the material within. In turn, we expect a
change in the orbital elements of the binary.

To estimate theoretically the magnitude of the back--reaction
a circumstellar disk has on the orbital elements of the binary
we assume an idealised system consisting of a binary system
and a ring-like mass distribution
orbiting star 1 with mass $m_{ring}$, at a distance ($\delta$-function)
of $r_{ring}$. The energy $E_{bin}$ and angular momentum $L_{bin}$
of the binary is given by
\begin{equation}
\label{eqn:el-bin}
   E_{bin} = - \, \frac{G M \mu}{2 a_{bin}}, \quad
   L_{bin} = \mu \left( G M a_{bin} \, ( 1 - e_{bin}^2) \right)^{1/2},
\end{equation}
and the corresponding quantities of the ring are
\begin{equation}
   E_{ring} = - \, \frac{G M_1 m_{disk}}{2 r_{ring}}, \quad
   L_{ring} = m_{ring} \, \left( G M_1  r_{ring} \right)^{1/2},
\end{equation}
where $M = M_1 + M_2$ is the total mass of the two stars and
$\mu = M_1 M_2 / M$ is the reduced mass.
Now, suppose that the ring is shifted from its initial position
$r_{ring}^\alpha$ to a smaller radius $r_{ring}^\beta$ keeping all
its mass. This radius change mimics the truncation of disk by the
binary. Through this process the ring's energy and angular momentum are
reduced from $E_{ring}^\alpha$ and $L_{ring}^\alpha$ to
$E_{ring}^\beta$ and $L_{ring}^\beta$.
By conservation of total energy and angular momentum
\begin{equation}
   E = E_{ring}  + E_{bin} \quad
   L = L_{ring}  + L_{bin}.
\end{equation}
We can calculate the corresponding change in the orbital elements
of the binary from $E_{bin}^\alpha$ and $L_{bin}^\alpha$ to
$E_{bin}^\beta$ and $L_{bin}^\beta$.
For the binary parameter masses $M_1 = 1.6 M_\odot, M_2 = 0.4 M_\odot$ with initial
orbital elements $a_{bin}^\alpha =18.5$AU and $e_{bin}^\alpha=0.36$
we find for the shift of a ring with
 $m_{ring}= 4 \times 10^{-3} M_\odot$  and initial radius $r_{ring}^\alpha = 4.0$ AU
to a final radius of $r_{ring}^\beta = 2.0$ AU that the binary elements change
to $a_{bin}^\beta =19.4$ AU and $e_{bin}^\beta=0.41$.
A quite substantial change considering the smallness of the ring's mass
in comparison to the stellar masses. But the closeness of the ring to the which is primary
allows it to gain a substantial amount of binding energy.
The calculation is approximate in the sense that the energy
and angular momentum of the ring are calculated with respect to
the non-inertial coordinate frame centred on the primary.

\begin{figure}[ht]
\def\capfrac{1}
\resizebox{0.99\linewidth}{!}{%
\includegraphics{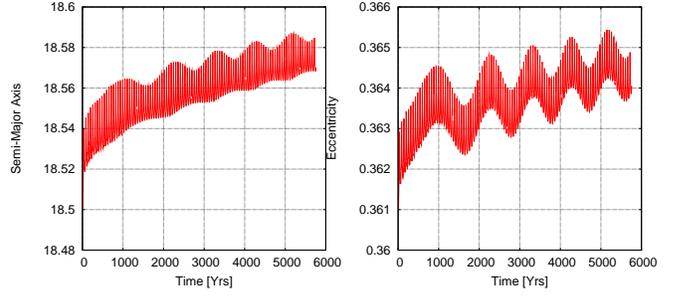}}
  \caption{
The evolution of the binary elements due to the interaction with the circumstellar
disk around the primary star, without an embedded planet.
One binary orbit refers to approximately 57yrs.
{\bf Left}: $a_{bin}(t)$; {\bf Right}: $e_{bin}(t)$.
  }
   \label{fig:aeb-h03}
\end{figure}
We can now compare this estimate with the previous hydrodynamical simulations,
and plot in Fig.~\ref{fig:aeb-h03} the evolution of $a_{bin}$ and $e_{bin}$
for about the first 100 binary periods without a planet included.
As demonstrated above, the binary expands as it gains energy from the truncated disk
and increases its eccentricity. The increase in $e_{bin}$ does not lead to
a decrease in the angular momentum however, since it increases its separation, see
Eq.~\ref{eqn:el-bin}.
Whenever the binary is near periastron the gravitational interaction with the
disk is maximal which results in the strong periodic spikes in the binary elements.
The change in the orbital elements of the binary is somewhat smaller than
the estimated values because {\it i}) the disk mass is smaller in the hydrodynamic
calculation and {\it ii}) disk mass and angular momentum are stripped off
by the secondary and are lost through the outer boundary of the computational domain.
The loss through the (open) inner boundary of the disk is only marginal.

\section{Evolution with an embedded planet}
\label{with-planet}
In the previous section we have analysed the change of the orbital elements
of the binary caused by the mass rearrangement in the disk.
Now we turn to the embedded planet in the disk, which is inserted into
the disk after an equilibration time of 100 binary orbits (nearly 6000 yrs).
This rather time consuming procedure to generate the initial
state is necessary to obtain realistic initial conditions for the
growing protoplanet.
At the time of insertion of the planet the remaining disk mass is rescaled
(through $\Sigma_0$) to contain 3 $M_{Jup}$ within the computational domain,
while keeping its two-dimensional distribution unchanged. 
To follow the growth and evolution from planetary cores to massive planets
the mass of the inserted protoplanet is $36 M_{Earth}$ orbiting the primary
star. 
This mass was chosen as it corresponds closely to the mass
above which forming gas giant planets undergo the final phase
of rapid gas accretion \citep{2005A&A...433..247P}.
The typical starting values of the semi-major axis and eccentricity are
$a_p = 2.5$AU and $e_p=0.0$.
\subsection{Exploratory models and numerical tests}
\label{subsec:exploratory}
After inserting the protoplanet on a circular orbit at 2.5 AU we expect that
its orbital elements will change due to the gravitational influence of the
binary and the distorted disk.
To differentiate the different contributions we decided to check
the origin of the dynamical behaviour, through a variation of physical conditions.
The standard model resembles the true physical situation where the
planet feels the full influence of the binary and the disk which is perturbed
by the binary. In the other set--ups we switch the various contributions on and
off. Avoiding additional complications in this analysis, the protoplanet
is not allowed to accrete here ($f_{acc}=0$) and its mass remains fixed at
$36 M_{Earth}$.

\begin{figure}
 \centering
 \includegraphics[width=0.95\linewidth]{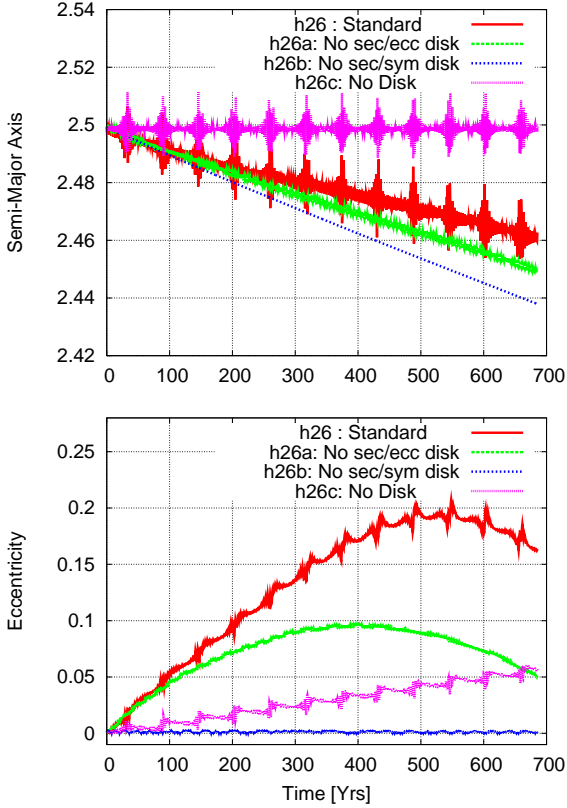}
 \caption{Time evolution of the semi-major axis and
    eccentricity of the embedded planet for different physical set--ups, neglecting
    either the secondary (h26a,h26b) or the disk (h26c), see text for details.
   \label{fig:aep-h26b}
   }
\end{figure}

The results are displayed in Fig.~\ref{fig:aep-h26b}, where the
semi-major axis and eccentricity of the planet are shown for four
different set--ups.
The standard model (h26, dark solid line) refers to the full model (including
binary and disk) as just described, in model
(h26a, dashed bright line) the mass of the secondary has been switched off to
test its influence, and the planet evolves in the initially eccentric disk
which becomes more circular during the evolution because of the absence
of the secondary.
In model (h26b, dark dotted line) in addition to having no secondary the density in the disk
has been azimuthally symmetrized keeping the radial distribution intact. Hence,
this model suits as a reference of what happens in the single star case.
In the last model (h26c, fine dotted line) the secondary is present but the disk
mass has been reduced, such that effectively only a 3-body problem is solved.
The curves with the periodic bumps in  Fig.~\ref{fig:aep-h26b} refer to the cases
(h26,h26c) including the secondary while the smoother curves describe the situation
where the secondary star has been excluded (h26a,h26b). 
The bumps occur with the binary period and indicate the `kick' the planet
experiences due to the interaction with the spiral arms in the disk and the
direct interaction with the binary companion when it is at periastron.
The results show that the main contribution to the {\it initial} growth of planetary
eccentricity $e_p$ is in fact the eccentric disk. The eccentricity $e_p$ for models
(h26 and h26a) rises initially with the same speed but due to the fact that, without the
secondary star, the disk slowly circularises, the model (h26a) falls behind the full
model (which reaches $e_p=0.2$ at $t\approx 500$) and $e_p$ begins to drop off after
a time of about 400 yrs after insertion of the protoplanet.
The maximum eccentricity reached for this reduced case (with no binary) is only $e_p=0.1$.
In the symmetrized disk case without secondary (h26b) the
planet behaves as expected for single star disks, it displays inward migration
(the fastest for all cases) and the eccentricity remains zero.
The case with only the binary does not show any change in the semi-major axis $a_p$
but rather a slow rise in $e_p$.
Additional test cases with similar set--up but different starting distances $a_p=3.0$ and
3.5 AU display similar behaviour. For the largest initial distance 3.5 AU there
are clear jumps in $e_p$ evolution visible at each individual periapse.
Hence, the rise of planetary eccentricity above the disk eccentricity for the full
model is a combination of disk and binary effects. 

\begin{figure}
 \centering
 \includegraphics[width=0.95\linewidth]{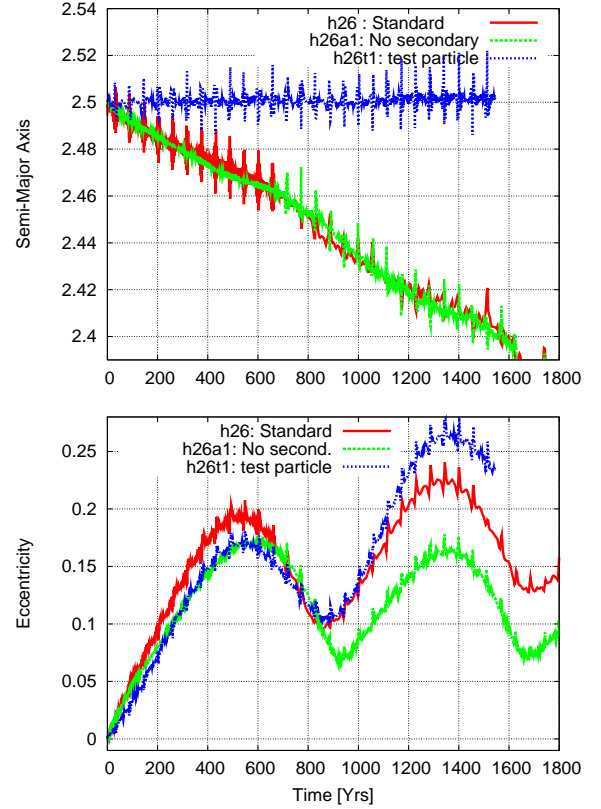}
 \caption{Time evolution of the semi-major axis and
    eccentricity of the embedded planet for different physical set--ups. Here, 
    the standard case (h26), a modified model where the dynamical influence of the
    secondary on the planet has been switched off (h26a1), and a model (h26t1) with a highly
    reduced planet mass (test particle with $5 \times 10^{-15} M_{Earth}$) are displayed.
   \label{fig:aep-h26t1}
   }
\end{figure}
\begin{figure}
 \centering
 \includegraphics[width=0.95\linewidth]{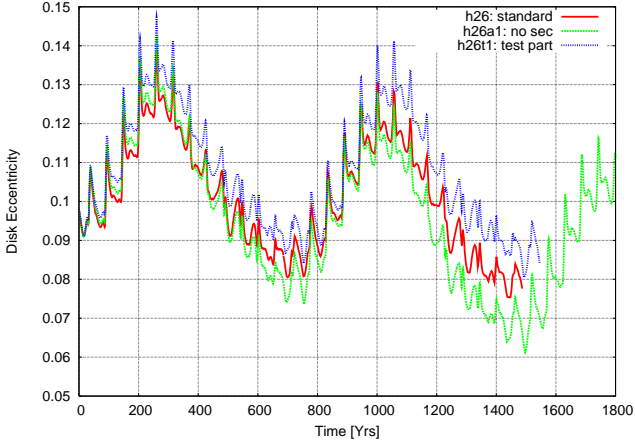}
 \caption{Time evolution of the averaged disk eccentricity for three
    different physical set--ups. 
    The standard case (h26), a modified model where the dynamical influence of the
    secondary on the planet has been switched off (h26a1), and a model (h26t1) with a highly
    reduced planet mass (test particle with $5 \times 10^{-15} M_{Earth}$) are displayed.
   \label{fig:ed-h26t1}
   }
\end{figure}
To further study the consequences of the individual contributions we have displayed in
Fig.~\ref{fig:aep-h26t1} in addition to the standard model a model (h26a1) where just the dynamical
(gravitational) effect of the secondary on the planet has been taken out. In
contrast to the previous model (h26a) the secondary still acts here on the disk. In addition
we display a model (h26t1) with a highly reduced planet mass of $5 \times 10^{-15} M_{Earth}$.
The migration rate of the first two models is very similar while reduced (nearly zero)
for the latter due to the smaller mass.
The eccentricity evolution is initially very similar for all 3 models, while at later
times the test particle simulation displays the largest planetary eccentricity growth and the model
without the secondary the smallest. 
We suspect that the faster growth of $e_{planet}(t)$ for the test particle is a result
of the vanishing migration and consequently larger influence of the binary.
In an additional plot (Fig.~\ref{fig:ed-h26t1})
the evolution of the averaged disk eccentricity for these models is displayed.
In the very small test particle case (model h26t1) the disk eccentricity is only influenced by the
binary and the $e_{disk}(t)$ evolution is a direct continuation of the initialisation process
(refer to Fig.~\ref{fig:ecc-om-h03}). 
From Fig.~\ref{fig:aep-h26t1} and \ref{fig:ed-h26t1} 
it is apparent that there
exists a phaseshift between the eccentricity of the disk and that of the planet.
As demonstrated above (cf. Fig.~\ref{fig:ecc-om-h03}) the 
variation of $e_{disk}(t)$
is correlated with the precession rate of the disk,
and is due largely to the secular disk--binary interaction.
The planetary eccentricity on the other hand is driven primarily by 
secular interaction with the disk (as shown above), and the phaseshift
between disk and planet eccentricity is caused by the shift in
disk and planet longitudes of pericentre induced by their relative precession
(see Fig.~\ref{fig:omdp-h28b}).

The two other models demonstrate that
the disk eccentricity is clearly influenced by the presence of even a small mass planet
of $36 M_{Earth}$. In the case where the influence of the secondary has been switched
off the planet's eccentricity is smaller as is the disk eccentricity. These results
demonstrate that the combined evolution of planet and disk properties are rather
intricately related in the case of binary stars. Hence, the evolution of the planet in this situation
is more complicated in contrast to a planet embedded in an eccentric disk without
a secondary star \citep[see][]{2002A&A...388..615P}, and not directly comparable.

\begin{figure}
 \centering
 \includegraphics[width=0.95\linewidth]{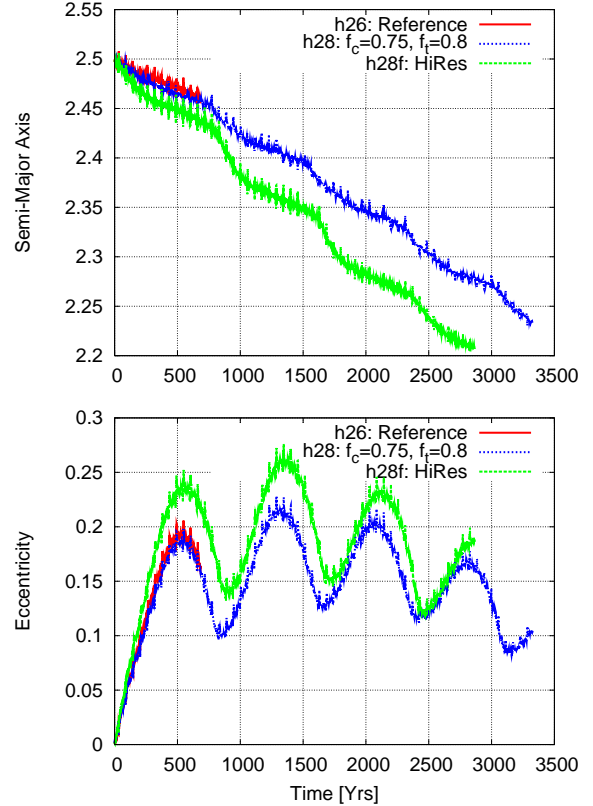}
 \caption{Time evolution of the semi-major axis and
    eccentricity of the embedded planet for different numerical set--ups. For the
    reference case (h26) we used for the Courant number and force cutoff $f_c = 0.5$
    and $r_t = 1.0$, respectively. The second model (h28: dark dashed line, run only until
    $t \approx 700$yrs) used $f_c = 0.75, r_t = 0.8$,
    which has also been used for the last model (h28f: light dashed)
     that has a grid resolution of $500\times500$.
   \label{fig:aep-h28f}
   }
\end{figure}
To test the influence of numerical issues we perform for the reference case (h26)
additional simulations with different set--ups. Among the purely numerical parameters
altered are: the time step length as given through the Courant number $f_c$, the force
cutoff radius $r_t$ as described in Sect.~\ref{subsec:numerical}, and the grid
resolution. The reference case in this section has been calculated with $f_c = 0.5$,
$r_t = 1.0$ and a resolution of $300\times300$.
As displayed in Fig.~\ref{fig:aep-h28f}, a change in the time--step and force cutoff 
to $f_c = 0.75, r_t = 0.8$ (dark dotted curve until $t=3300$) has
only a marginal influence. A change of resolution, here to $N_r \times N_\varphi = 500\times 500$
leads to a faster migration rate (30-40\%) and slightly larger eccentricities.
The qualitative behaviour (inward migration and saturation of eccentricity) is not changed,
however. Given these encouraging results, we use for computational reasons the reference resolution
of $300\times300$ and $f_c = 0.75, r_t = 0.8$ in our subsequent models.
\subsection{Models without accretion onto planet}
\begin{figure}
 \centering
 \includegraphics[width=0.95\linewidth]{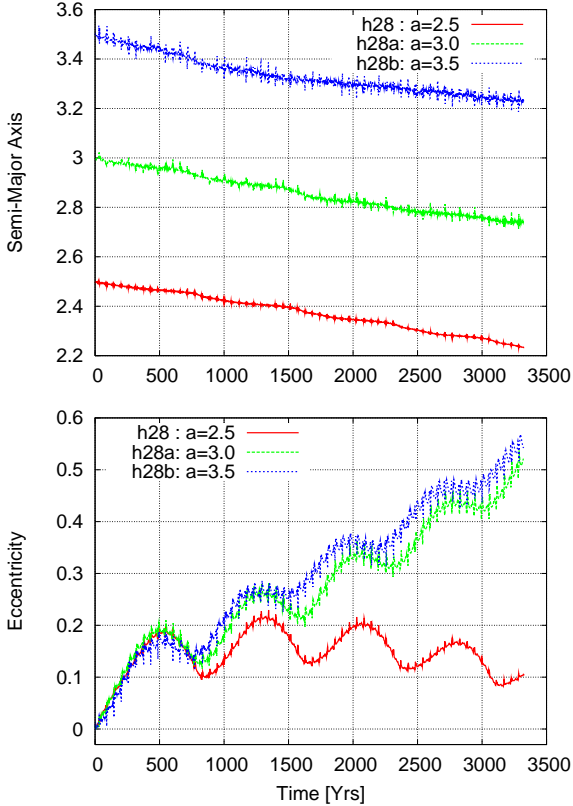}
 \caption{Time evolution of the semi-major axis and
     eccentricity of a non-accreting planet embedded planet for three
     different starting locations.
   \label{fig:aep-h28b}
   }
\end{figure}

Planetary cores form in the outer cooler regions of protoplanetary disks beyond
the so--called ice-line. In a binary star system the outer disk is affected most
by the secondary, and to find possible restrictions on the planet forming regions
in the disk it is important to analyse the evolution of cores near
the outer parts of the disk.
To study the effect of initial position we start our embryos at 3 different locations
in the disk 2.5, 3.0 and 3.5 AU always on a circular orbit, and choose again
non-accreting cores. Because the initial characteristic growth time of the cores may be long,
even in comparison to the orbital period of the binary
these sets of runs constitute a test suite to estimate the orbital evolution
of protoplanets in the disk. 
The results for the semi-major axis and eccentricity evolution of the planet
are displayed in Fig.~\ref{fig:aep-h28b}, where the only difference in the three
cases is the release distance (2.5, 3.0 and 3.5 AU) of the planet.
From all three locations the planets migrate inwards at approximately the same rate
with the tendency for a slowdown for the two outer cases. This inward migration
of the outer planets which are mainly orbiting outside the disk is due the
the planet disk interaction near planetary periastron where the planet moves
faster through the disk and induces a negative torque and power (energy loss)
\citep{2007A&A...473..329C}.
However, the different initial starting radii lead to a very different eccentricity
evolution.  While only the innermost planet (starting at 2.5 AU) shows a finite
eccentricity evolution
the two outer cases display a very strong increase in their eccentricity beyond
$e_p=0.5$ after about 55 binary orbits.
Clearly the strongly disturbed disk in the outer regions at around 4~AU significantly
perturbs the orbits of the protoplanet and does not allow for small equilibrium
eccentricity.
While the initial faster rise of $e_p$ 
appears to be caused primarily by the disk,
the long--term evolution to very high eccentricities is most likely due to the
combined action of the binary and the eccentric disk.
A test simulation (for an initial semi-major axis $a_p=3.0$~AU)
where the gravitational action on the planet due to the secondary 
has been switched off did not display any large eccentricity growth but a steady
decline. On the other hand, starting the planets with high eccentricity 
(from $a_p=3.0$~AU) in a pure 3-body problem does also not lead necessarily to
a growth of eccentricity but rather to an oscillatory behaviour.
So we may conclude that in the present situation it is the combined
action of an eccentric, periodically perturbed disk {\it and}
the influence of an eccentric binary star.
\begin{figure}
 \centering
 \includegraphics[width=0.85\linewidth]{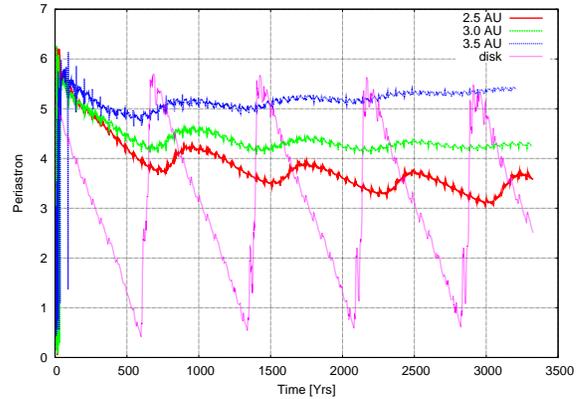}
 \caption{Time evolution of the argument of pericentre (periastron) for
     an embedded planet starting at 3 different locations and the protoplanetary
     disk. 
   \label{fig:omdp-h28b}
   }
\end{figure}

As the planets move on non-circular orbits in an eccentric disk and binary 
a temporal change of the apsidal line may be expected.
In Fig.~\ref{fig:omdp-h28b} the evolution of the argument of pericentre of the
planets and the disk are plotted versus time. The innermost planet has on average
a periastron angle of about $200\deg$ while the outer planets have a larger
angle. For the binary the angle of periastron lies fixed at 0 degrees, and the disk
is slowly precessing retrograde as pointed out above and indicated in
Fig.~\ref{fig:omdp-h28b} by the tightly dashed line. The influence of the disk
on the planets is visible by the slight oscillations of the periastron angle
about the mean with the same period as the oscillations in the eccentricity.
The innermost planet has approximately a phase shift of $180\deg$ with respect to
the binary and is nearly in an anti-symmetric state while the other planets are
lagging behind this configuration.
\subsection{Models with mass accretion onto planet}
\begin{figure}
 \centering
 \includegraphics[width=0.95\linewidth]{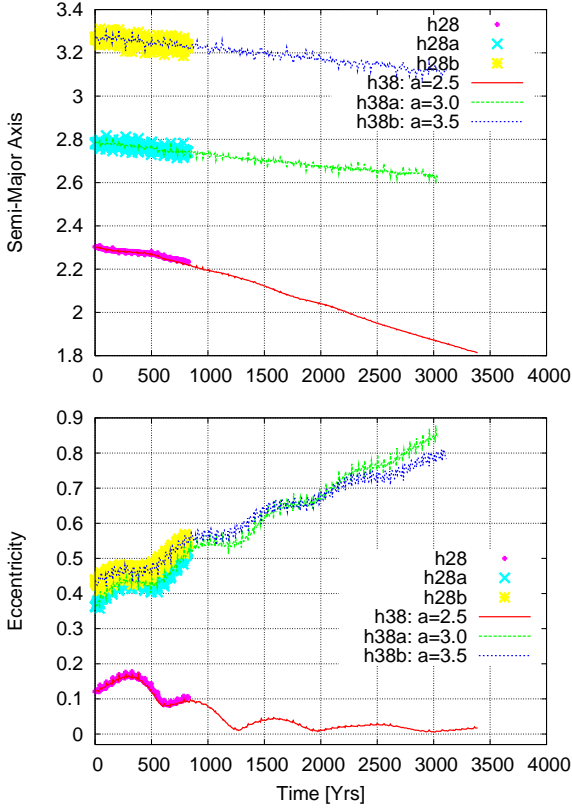}
 \caption{Time evolution of the semi-major axis and eccentricity for 3 {\it accreting}
   protoplanets starting at 3 different locations in the disk. The symbols refer to the
   non-accreting models of Fig.~\ref{fig:aep-h28b} time--shifted by 2500 yrs, while the lines
   refer to restarted models with accretion switched on.
   \label{fig:aepc-h38a}
   }
\end{figure}

To make the scenario more realistic we add now the option of mass
accretion onto the protoplanet. To do so, we assume that initially the planets grew
during a long time to their present mass and enter now the more rapid gas accretion
phase. Numerically we continue the previous simulations for the
three initial radii at a time $t=2500$\,yrs. The evolution of the orbital elements
of the planets are displayed in Fig.~\ref{fig:aepc-h38a} where the time offset has been
reset after the turn-on of mass accretion.
The thinner lines (until $t\approx 3000$) are the models with accretion switched on
at a low rate using an efficiency of $f_{acc}=3 \times 10^{-2}$. The symbols refer
to the final phase of the models without accretion.
While the eccentricities of
the outer planets continue to rise to very high values despite a slow inward migration,
the inner planet's orbit (starting at $a_p = 2.3$AU) begins to circularise while
the inward migration speed increases.
The large eccentricities of the outer planets can probably
only be maintained in a stable way because of some eccentricity damping action of the disk
during their periastron passage.
\begin{figure}
 \centering
 \includegraphics[width=0.85\linewidth]{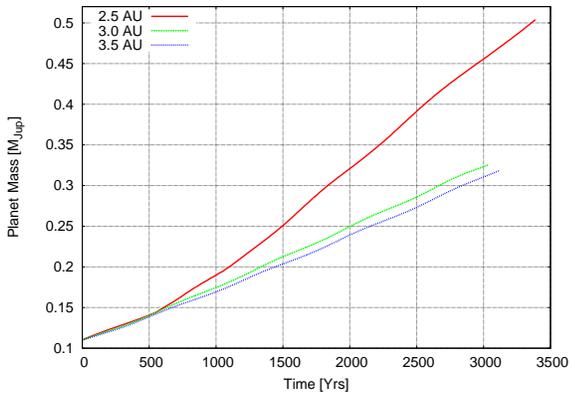}
 \caption{Mass of the planet vs. time for accreting models for 3 different initial locations
  of the planet.
   \label{fig:mp-h38a}
   }
\end{figure}
The planetary mass evolution for the three cases is displayed in  Fig.~\ref{fig:mp-h38a}.
The innermost planet has reached about half a Jupiter mass after 3400 yrs while the other
two grow at a smaller rate and have reached about 0.3 $M_{Jup}$ after 3000 yrs.
Note that these labels refer to the initial positions of the planets that were first evolved
without mass accretion and then continued with $f_{acc} \neq 0$ as described above.
From the simulations it seems to be clear that it may be difficult to grow a planet
at too large a distance from the primary star. For our chosen binary parameter the borderline
lies around $2.7$ AU, which has been verified by 
additional simulations with different starting distances
that are not shown here. This particular transition radius of about $2.7$ AU
depends on the chosen parameters for the disk viscosity and temperature.
Larger values can possibly extend this transition radius slightly while smaller
values will reduce it. 
Varying the disk mass will change the viscosity and temperature of the disk,
and we expect this to influence the transition radius, too.
It is beyond the scope of the present to paper to perform
detailed parameter studies on these issues.
From the results displayed in Fig.~\ref{fig:mp-h38a} it appears
that mass accretion onto the planet does not alter the orbital evolution of the
planet significantly.

\begin{figure}
 \centering
 \includegraphics[width=0.95\linewidth]{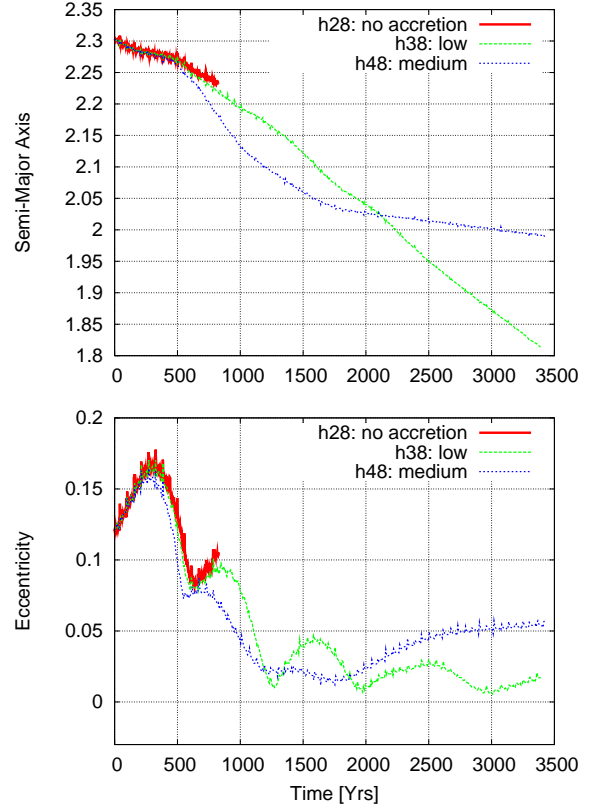}
 \caption{Time evolution of the semi-major axis and eccentricity of 3 accreting protoplanets
   starting at the same initial condition with different mass accretion rates.
   \label{fig:aepc-h48}
   }
\end{figure}

To further explore parameter space we used the starting condition 
$a_p=2.5$ and performed one additional run with a three times higher accretion rate onto the
planet ($f_{acc} = 10^{-1}$, denoted `medium' in the figures).
The results of this new model with the previous
no accretion and low accretion models are shown in Fig.~\ref{fig:aepc-h48}.
\begin{figure}
 \centering
 \includegraphics[width=0.85\linewidth]{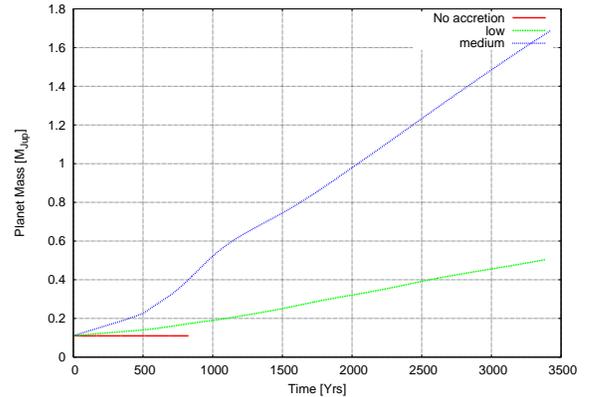}
 \caption{Mass of the planet vs. time for accreting models starting at $a_p=2.5$AU for
  3 different accretion rates.
   \label{fig:mp-h48}
   }
\end{figure}
The results for the orbital elements of the planet 
in Fig.~\ref{fig:aepc-h48} 
show that the models with no and low accretion behave similarly.
The planet continues to migrate inward while decreasing its eccentricity.
The model with the higher accretion rate migrates faster at first, and slows down
near the end, while the eccentricity settles to about $e_p = 0.05$.
This reduction in the migration rate is a consequence of the increasing mass of the
planet. For this higher accretion rate the planet's mass increases to nearly
$1.7 M_{jup}$ after 3300 years (see Fig.~\ref{fig:mp-h48}).
Since the disk mass is reduced by the same amount the
driving agent of the planet is lost and the speed of migration reduced.
This model with the `medium' accretion rate settles to a final state which is quite
close to the observed value for the $\gamma$ Cep system.

\begin{figure}[ht]
 \centering
 \includegraphics[width=0.95\linewidth]{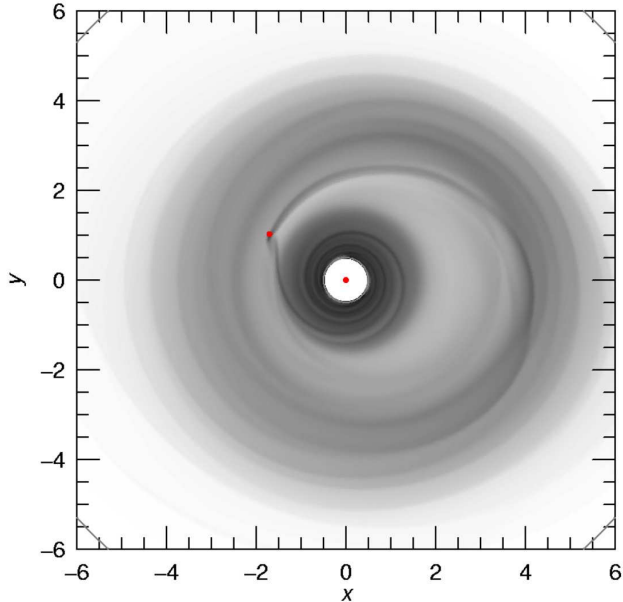}
 \caption{Grayscale plot of the two dimensional density distribution 
  of the medium accretion model (h48) at time $3125$~yrs. 
 The shading is scaled $\propto \Sigma^{1/4}$ between $4.8 \times 10^{-4}$ (white) 
  and $2400$ g/cm$^{2}$ (black). 
  The location of the planet
  is marked by the small circle.
   \label{fig:h48-sig2d}
   }
\end{figure}

\begin{figure}[ht]
\def\capfrac{1}
\resizebox{0.47\linewidth}{!}{%
\includegraphics{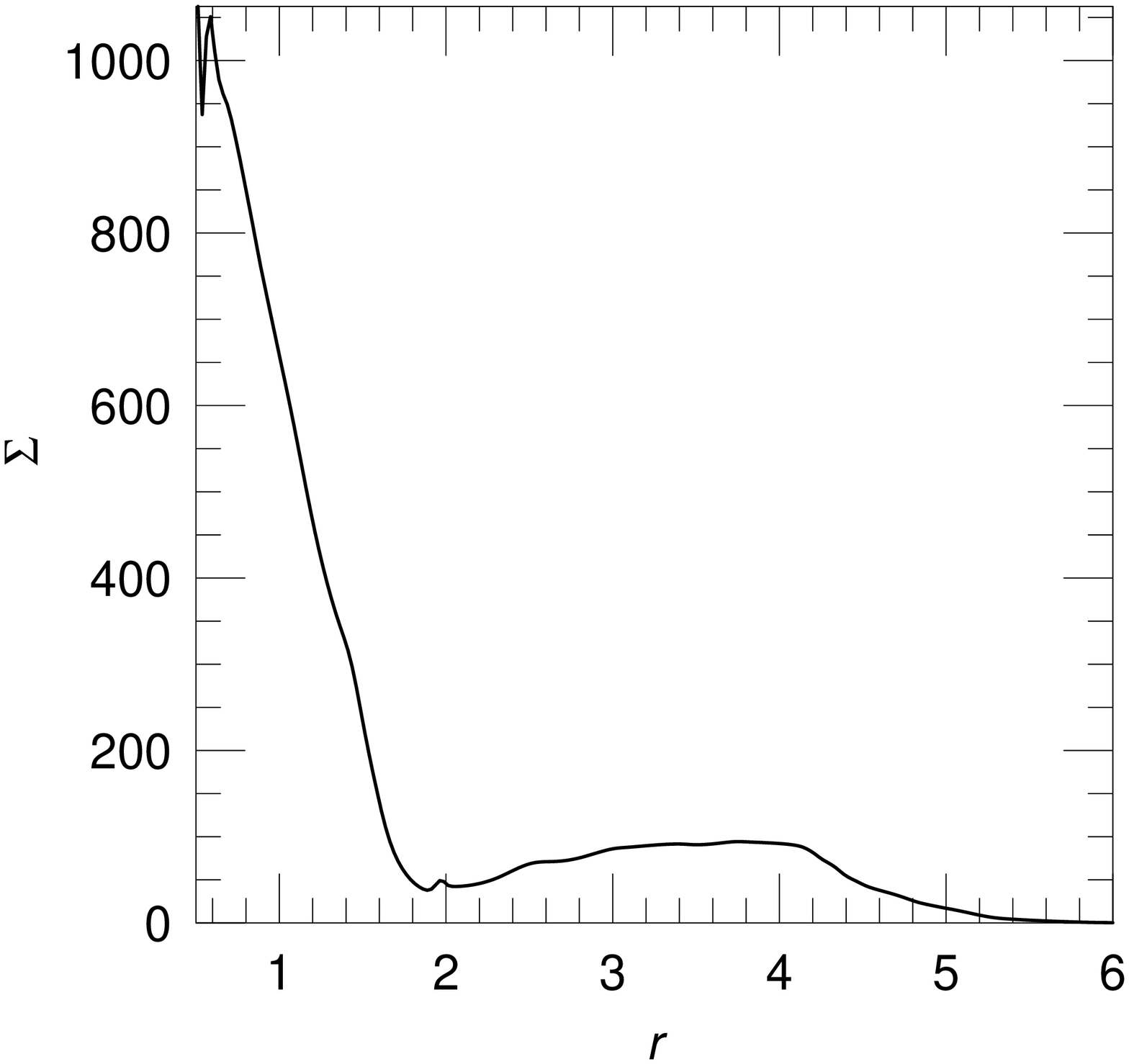}}
\resizebox{0.47\linewidth}{!}{%
\includegraphics{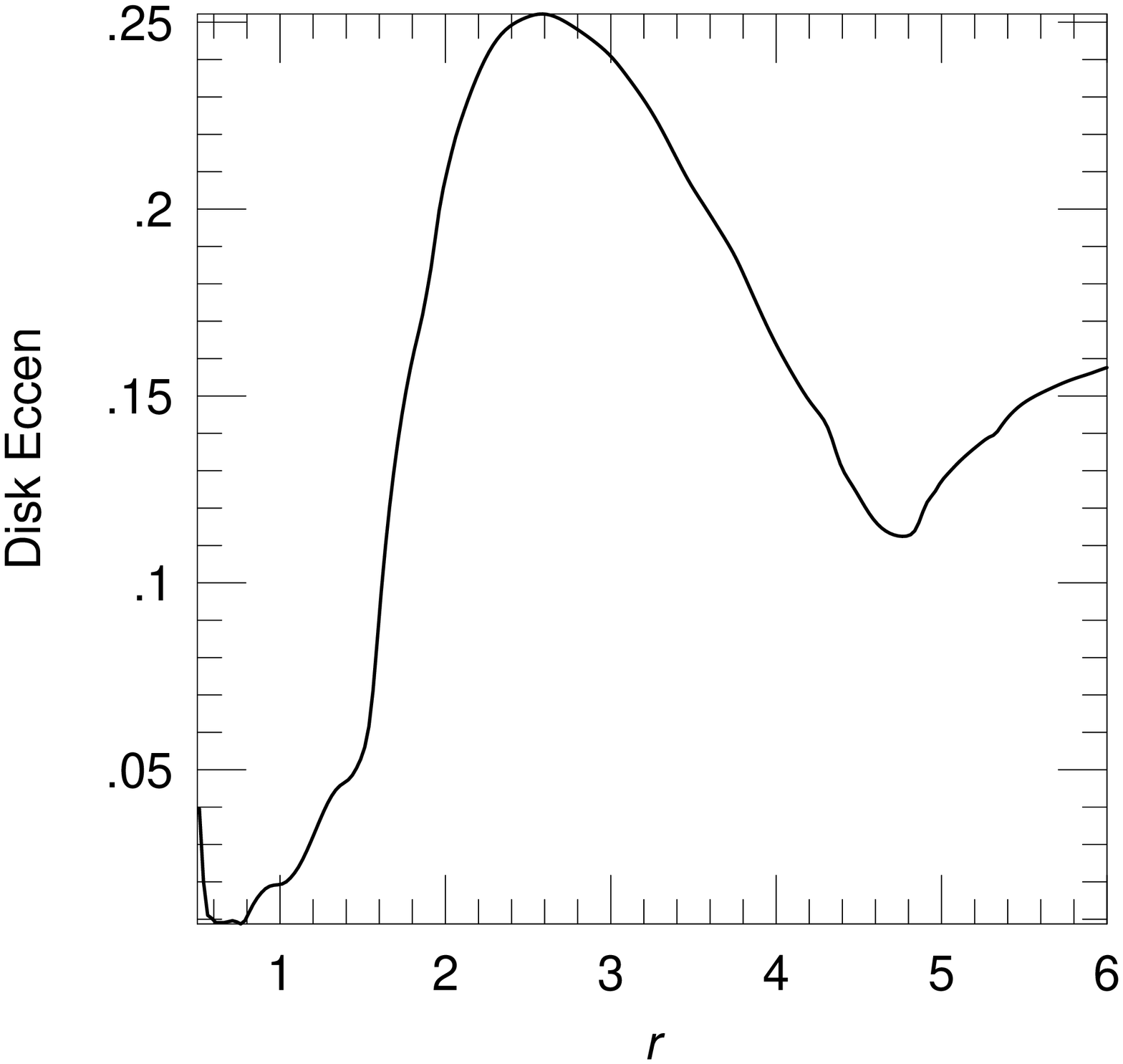}}
  \caption{
Azimuthally averaged density (in g/cm$^{2}$) and eccentricity of the disk for the medium accretion
model at a time of $3125$~yrs.
  }
   \label{fig:h48-sig-ecc}
\end{figure}
A massive embedded planet will open a gap in standard circular disks, and it
is interesting to analyse this effect within the present context. In Fig.~\ref{fig:h48-sig2d}
we display the two-dimensional density distribution $\Sigma(r,\varphi)$ in the disk
at a time $3125$~yrs for the medium accretion model. At this time the planet has reached
approximately a mass of 1.5 $M_{Jup}$. From the plot it seems that the disk
inside the planetary orbit is apparently more circular than outside.
This is confirmed by the corresponding one-dimensional radial distribution of
the azimuthally averaged density and eccentricity of the disk at the same time.
The gap is somewhat weaker than in circular disks primarily due to the periodic
disturbance of the secondary that tends to sweep material into the cleared region
around the planet. 
Due to the shallower gap the planet is able to continue mass accretion
from its surroundings more easily compared to a planet on a circular
orbit in a single star system.
The inner disk clearly has a smaller eccentricity than the outer parts
(compare this to Fig.~\ref{fig:sig-ecc-h03}). The presence of the planet represents
a barrier for the (spiral) wave induced by the binary which consequently cannot propagate
into the inner parts of the disk.

\section{Conclusions and Summary}
\label{conclusion}
In the present work we have concentrated on the planetary formation in the
system $\gamma$~Cep which places, due to its binary parameter
of $a_{bin} = 20$AU and $e_{bin}=0.41$, 
severe constraints on the formation process.
We investigated an intermediate  phase in the planet formation process
within the so-called core formation scenario, i.e.
we did not attempt to model the formation of planetesimals in this system but
rather concentrated on a later phase when protoplanetary cores have already
formed and begin their rapid gas accretion phase. 

Before embedding the protoplanets in the disk, we have first brought the system
into equilibrium by performing simulations with no embedded protoplanets.
This initialisation phase takes about 100 binary orbits after which the
disk structure has settled to a new quasi-equilibrium configuration with a truncated
disk. Interestingly, the accretion disk around the primary reaches an average
eccentricity of about $e_{disk} = 0.12$ and shows a coherent {\it retrograde}
precession, in agreement with the findings of
\citet{2008arXiv0802.0927P}. A detailed parameter study of this interesting behaviour
on its own is beyond the present analysis and will be presented elsewhere.
During one binary orbit the eccentric secondary disturbs the
disk periodically and induces significant non-axisymmetric perturbations
in the disk which decay due to viscosity during binary apoapse.

As suspected these perturbations of the disk, in particular the periodic creation
of strong tidally induced spiral density waves and the creation of
an eccentric disk, lead to non-negligible effect on
the planetary orbital elements. While embryos placed in the disk at 
different initial distances
from the primary star continue to migrate inward at approximately the same rate, the
eccentricity evolution is markedly different for the individual cases.
If the initial distance is beyond about $a \gsim 2.7$ AU the eccentricity of the embryo
continues to rise to very high values and, apparently only due to the damping action of the
disk, the orbits remains bound. The excitation mechanism of the eccentricity is
the combined action of the binary and the perturbed disk (i.e. there is a
long term secular interaction due to the disk being eccentric).
Test simulations where individual components of the system have been
switched off and on have made it clear that it is indeed the combination of effects
(eccentric binary and the eccentric, periodically perturbed disk) that leads
to the high planetary eccentricities in the system.
Once $e_{planet}$ has grown to sufficiently large values the planetary orbit
may reach the stability limit as given for example by the pure 
3-body simulations for $\gamma$ Cep \citep[see Fig.~2 in][]{2004MSAIS...5..127T}.

A low mass, non accreting planet embedded in an
the eccentric disk experienced substantial growth in eccentricity
(see Fig.~\ref{fig:aep-h28b}).
This has clear implications for the accretion of planetesimals
because their velocity dispersion may become very large due to this
effect. \citet{2004A&A...427.1097T} examined the evolution of
planetesimal orbits under the influence of the binary companion
and aerodynamical gas drag. They concluded that accretion of
planetesimals would occur in the shear dominated regime
because orbital alignment was maintained due to the gas drag.
This work, however, did not include the effects of an eccentric
disk, and so it remains unclear whether planetesimal orbits will
remain aligned. Recently, this issue has been addressed by
\citet{2007arXiv0705.3421K} and
\citet{2008arXiv0802.0927P} who find that the inclusion of the dynamical evolution
of the disk may render the planetesimal formation more difficult, but this
topic certainly deserves more consideration.

The most up to date observational data suggest the following
parameters for the planet in the $\gamma$ Cep system:
$a_p \simeq 2.044$, $e_p \simeq 0.115$ and 
$m_p \sin{i} \simeq 1.60$ M$_{Jupiter}$. If this planet formed
according to the core instability model, 
then an important issue is the survival of the planetary core
before gas accretion occurs. Fig.~\ref{fig:aep-h28b}
has shown that the non-accreting, low mass planet undergoes quite rapid inward
migration. The migration, however, is modulated by the eccentricity
of the planet, such that at high eccentricity phases the migration rate
decreases. It is possible that longer run times will show an
essential stalling of this migration if the planet eccentricity
grows beyond its final value of $e_p \simeq 0.3$.
The inclusion of a better approximation to the physical state
of the disk, eg. radiative cooling, may help to slow down
this initial phase, as new results indicate that
in case of radiative disks migration is slowed down or even reversed
for protoplanetary cores \citep{2008A&A...478..245P, 2008ApJ...672.1054B}.

Once gas accretion is switched on, it is clear that a disk mass of
about 3 Jupiter masses, where the outer disk radius is tidally
truncated at $r \simeq 5$ AU, will be sufficient to grow a planet
that is close to the minimum observed mass of $m_p \sin{i} \simeq 2.044$
M$_{Jupiter}$. It is also clear that we can construct a model
in which a low mass planet growing
from an initially circular orbit can achieve a final mass of $m_p \simeq 2$
M$_{Jupiter}$, and have a final eccentricity of $e_p \simeq 0.1$ as required.

A final comment relates to the final mass of the planet.
Our simulations suggest that a disk mass of about 3 Jupiter masses
will be enough to form a gas giant of the required minimum mass.
A future test of the mode by which the planet in $\gamma$ Cep formed
(gravitational instability versus core accretion)
will be determination of its actual mass. We suspect that a disk
that is massive enough to form a planet through gravitational
instability will lead to a planet whose final mass is
substantially larger than the minimum value observed.

\begin{acknowledgements}
Very useful discussions with Dr. Paardekooper at the DAMTP in Cambridge
are gratefully acknowledged.
The work was sponsored in parts by the EC-RTN Network {\it The Origin of Planetary Systems}
under grant HPRN-CT-2002-00308, and by grant KL~650/6 of the German Research Foundation
(DFG). Some of the simulations conducted as part of this project were performed
on the QMUL High Performance Computing facility funded through the SRIF initiative.
\end{acknowledgements}
\bibliographystyle{aa}
%%\begin{thebibliography}
\bibliography{kley8}
\end{document}